\documentclass[aps,twocolumn,prb]{revtex4-2}
\usepackage{graphicx}
\usepackage{cases}
\usepackage{color}

\begin{document}

\title{Omnidirectional excitation of surface waves and super-Klein tunneling at the interface between two different bi-isotropic media}

\author{Seulong Kim}
\affiliation{Department of Energy Systems Research and Department of Physics, Ajou University, Suwon 16499, Korea}
\author{Kihong Kim}
\email{khkim@ajou.ac.kr}
\affiliation{Department of Energy Systems Research and Department of Physics, Ajou University, Suwon 16499, Korea}

\begin{abstract}
We study theoretically some unique characteristics of surface electromagnetic waves
excited at the interface between two different kinds of general bi-isotropic media,
which include Tellegen media and chiral media as special cases.
We derive an analytical dispersion relation for those waves,
using which we deduce eight different conditions under which they are generated between
two Tellegen media and between two chiral media independently of the component
of the wave vector along the interface. These make it possible to excite the surface waves
for all or a wide range of incident angles in attenuated total reflection experiments on multilayer structures.
We generalize the concept of a conjugate matched pair to bi-isotropic media and obtain several conditions under which
the omnidirectional total transmission, which we call the super-Klein tunneling, occurs through conjugate matched pairs
consisting of Tellegen media and of chiral media.
We find that these conditions are closely linked to those for the omnidirectional excitation of surface waves.
Using the invariant imbedding method, we perform extensive numerical calculations of the absorptance, the transmittance, and the spatial distribution of the electromagnetic fields
for circularly-polarized waves incident on bilayer structures
and confirm that the results agree perfectly with the analytical predictions.
\end{abstract}

\maketitle

\section{Introduction}
\label{sec1}

Surface electromagnetic waves of various kinds have been a focus of intensive research in recent decades \cite{polo,taka}.
It has been well known that when these waves are excited on the surface of a medium or at the interface between
two media, the electromagnetic fields close to the surface or
the interface are greatly enhanced. The high sensitivity of this enhancement to various parameters
and the resulting linear and nonlinear optical effects
have been successfully applied in developing efficient photonic devices and sensors \cite{mayer,petry,amen}.
Strong interests in the physics and applications of these waves have even created a new research area called plasmonics \cite{maier,mis}.

Surface plasma waves or surface plasmons, which are the simplest type of surface electromagnetic waves,
can be excited on the surface of metals by external electromagnetic radiation \cite{raether}.
More recently, there has been a growing interest in the different types of surface waves in other complex media such as
Dyakonov waves excited at the interface of anisotropic media \cite{taka2,nari}, optical Tamm plasmon polaritons excited
on the surface of a photonic crystal \cite{xue,8,abf}, and surface plasmon polaritons associated with chiral media \cite{12,nah}. Surface polaritons on the surface of negative index media \cite{9,doc}, surface waves due to a spatial inhomogeneity
near the surface of semiconductors \cite{10,shv}, surface waves in general bi-anisotropic media \cite{gal}, and the influence of optical nonlinearity
on surface plasmons \cite{ksp1,ksp2,ksp3} have also attracted some attention.

In a recent paper, we have presented a detailed study of the characteristics
of surface waves excited at the interface between a metal and a general bi-isotropic medium \cite{sw}.
Bi-isotropic media, which include Tellegen media and chiral media as special cases, are the most general form of linear isotropic media, where the electric displacement {\bf D} and the magnetic induction {\bf B} are linearly and isotropically related to both the electric field {\bf E} and the magnetic intensity {\bf H} \cite{1,2}. In cgs Gaussian units, the constitutive relations for harmonic waves in these media can be expressed as
\begin{eqnarray}
{\bf D}=\epsilon{\bf E}+a{\bf H},~~
{\bf B}=\mu{\bf H}+a^*{\bf E},
\label{eq:cr}
\end{eqnarray}
where $\epsilon$ is the dielectric permittivity and $\mu$ is the magnetic permeability.
The magnetoelectric parameter $a$ is written as
\begin{equation}
a=\chi+i\gamma,
\end{equation}
where $\chi$ is called the non-reciprocity (or Tellegen) parameter and $\gamma$ is called the chirality index.

In the present paper, we generalize the theory of Ref.~\cite{sw} to the case where the surface waves are excited at the interface between two different kinds of general bi-isotropic media.
We first derive analytically the generalized dispersion relation for surface waves.
Our main focus in this work is to derive the explicit conditions under which surface waves are
generated between
two Tellegen media and between two chiral media independently of the component
of the wave vector along the interface, starting from the dispersion relation.
When these conditions are satisfied, it is possible to excite the surface waves
for all or a wide range of incident angles in attenuated total reflection (ATR) experiments on multilayer structures.
We call this phenomenon the omnidirectional excitation of surface waves.
We confirm these predictions by calculating
the absorptance and the spatial distribution of the electromagnetic fields
for circularly-polarized waves incident on multilayer structures
using a generalized version of the invariant imbedding method (IIM) \cite{kly,kim1,kim3,kim4,15}.

We also find that the omnidirectional excitation of surface waves is intimately related to the phenomenon of
omnidirectional total transmission of waves
through a conjugate matched bilayer. In Ref.~\cite{alu}, the authors have considered the wave propagation through
a pair of slabs with the medium parameters
$\epsilon_1$ and $\mu_1$ and $\epsilon_2$ and $\mu_2$, respectively. It has been demonstrated that when the condition $\epsilon_1/\epsilon_2=\mu_1/\mu_2=-1$ is precisely
satisfied and the thicknesses of the two slabs are the same, waves incident on the bilayer
are totally transmitted regardless of the incident angle and the polarization. The pair of slabs satisfying the above conditions has been called
the conjugate matched pair. We generalize this concept to bi-isotropic media and show that the omnidirectional total transmission also occurs in such cases in the complete absence of dissipation. If there exists a small dissipation in the same system,
however, we find that a finite absorption due to the omnidirectional excitation of surface waves always arises.

Recently, the omnidirectional total transmission of electron waves through a scalar potential barrier has been found to occur in pseudospin-1 Dirac-type materials, when the electron energy is one-half the value of the potential \cite{shen,urban,fang0,bo1,kim_rip,kimd2}. In an analogy to the Klein tunneling occurring when electrons are incident normally on a potential barrier of an arbitrary shape in Dirac materials \cite{kats2,been,nicol,kimd1}, this phenomenon has been
termed the super-Klein tunneling.
From the viewpoint of the physics of wave propagation, the origins of the the super-Klein tunneling and the omnidirectional total transmission through a conjugate matched pair are quite similar, and therefore we will refer the latter as the super-Klein tunneling as well.
This phenomenon is also studied in detail by calculating the transmittance using the IIM.

The rest of this paper is organized as follows.
In Sec.~\ref{sec2} we derive an analytical dispersion relation for surface waves at the interface between two different bi-isotropic media.
Using the dispersion relation, we derive eight different conditions for omnidirectional excitation of surface waves in Sec.~\ref{sec22}.
We also specify the conditions under which the omnidirectional total transmission through a conjugate matched pair occurs.
In Sec.~\ref{sec3} we develop a generalized IIM for wave propagation in stratified bi-isotropic media.
In Sec.~\ref{sec4} we perform extensive numerical calculations using the IIM and compare the results with
the predictions of the dispersion relation.
In Sec.~\ref{sec5} we comment on the experimental feasibility and conclude the paper.

\section{Dispersion relation for surface waves}
\label{sec2}

\begin{figure}
\centering\includegraphics[width=8cm]{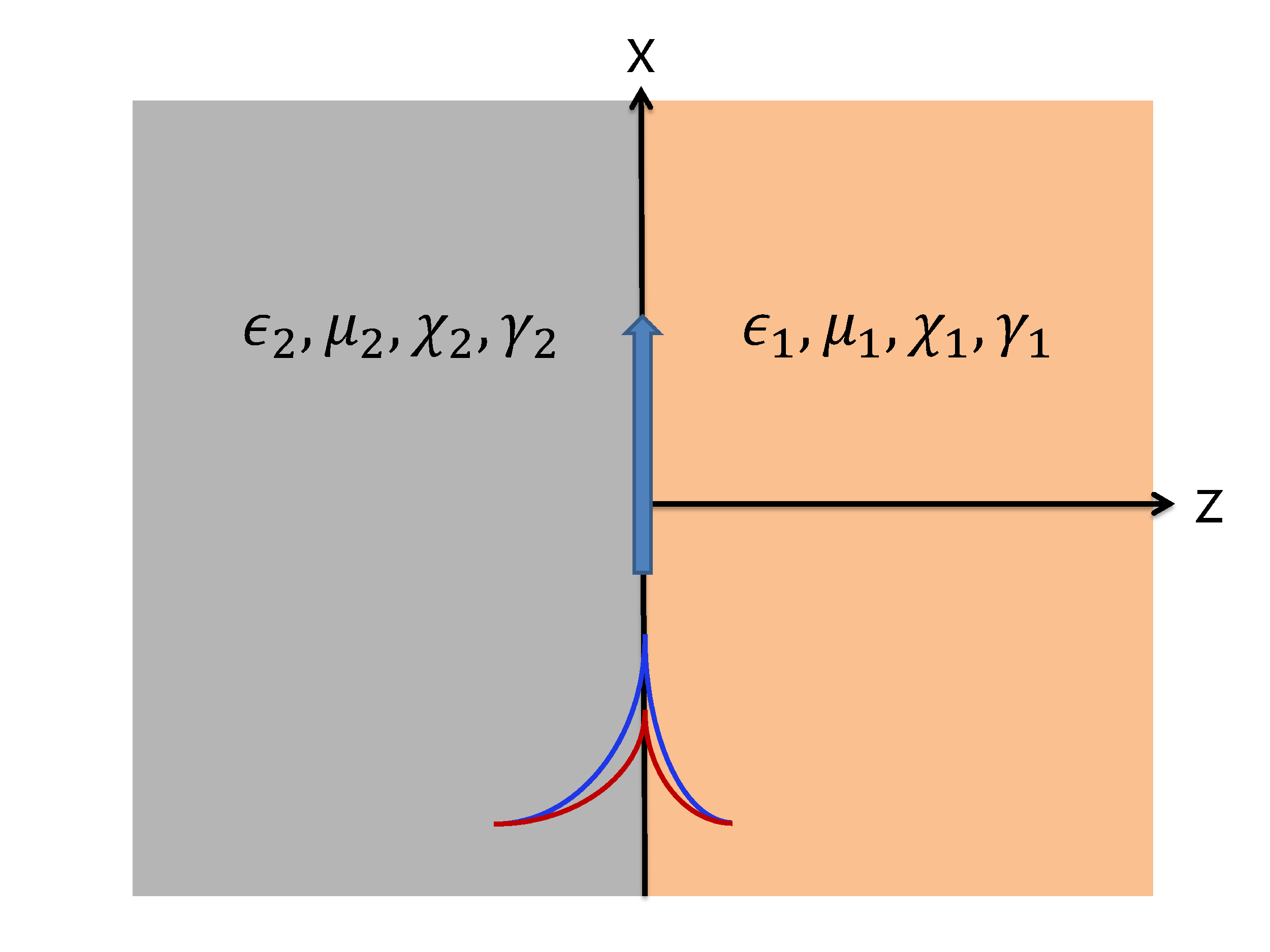}
\caption{Sketch of the surface wave propagating in the $x$ direction
along the interface at $z=0$ between two different kinds of general bi-isotropic media.
The blue and red curves illustrate the field amplitudes for different circular polarizations decaying exponentially away from the interface.
The decay rates, $\kappa_{jr}$ and $\kappa_{jl}$ ($j=1,2$), are different for RCP
and LCP modes and are related to the $x$ component of the wave vector $q$, which is a conserved quantity, by Eq.~(\ref{eq:kap}).}
\label{fig1}
\end{figure}

We consider a plane interface between two different kinds of general bi-isotropic media located at $z=0$ as illustrated in Fig.~\ref{fig1}.
We seek a surface-wave solution to Maxwell's equations that is propagating in the $x$ direction along the interface but
is exponentially damped away from it.
The two bi-isotropic media are characterized by the parameters $\epsilon_1$, $\mu_1$, and $a_1$ $(=\chi_1+i\gamma_1)$ and
$\epsilon_2$, $\mu_2$, and $a_2$ $(=\chi_2+i\gamma_2)$, respectively.
In a uniform bi-isotropic medium, there exist two eigenmodes of circular polarization and the electromagnetic fields can be decomposed as
\begin{eqnarray}
{\bf E}={\bf E}_r + {\bf E}_l,~~ {\bf H}={\bf H}_r + {\bf H}_l,
\end{eqnarray}
where the subscripts $r$ and $l$ denote right-circularly polarized (RCP)
and left-circularly polarized (LCP) modes respectively.
The effective refractive indices $n_{jr}$ and $n_{jl}$ and the effective impedances $\eta_{jr}$ and $\eta_{jl}$ for
RCP and LCP modes, respectively, in the bi-isotropic medium $j$ ($j=1,~2$) are given by \cite{1,15}
\begin{eqnarray}
&&n_{jr}=n_{j}+\gamma_j,~~
n_{jl}=n_{j}-\gamma_j,\nonumber\\
&&\eta_{jr}=\frac{n_{j}+i{\chi_j}}{\epsilon_j},~~
\eta_{jl}=\frac{n_{j}-i{\chi_j}}{\epsilon_j},
\label{eq:neta}
\end{eqnarray}
where $n_{j}$ is defined by
\begin{equation}
  n_{j}=\left\{ \begin{array}{lll}
         -\sqrt{\epsilon_j\mu_j-{\chi_j}^2}, & \mbox{if } \epsilon_j,\mu_j<0 ~\mbox{and}~ \epsilon_j\mu_j>{\chi_j}^2 \\
         i\sqrt{{\chi_j}^2-\epsilon_j\mu_j}, & \mbox{if } \epsilon_j\mu_j<{\chi_j}^2 \\
         \sqrt{\epsilon_j\mu_j-{\chi_j}^2}, & \mbox{otherwise}
       \end{array}\right..
\end{equation}
We note that there are cases where some of the effective refractive indices take negative values, such as in ordinary negative index media with
$\epsilon<0$, $\mu<0$, and $\chi=0$.

In the $z>0$ region, ${\bf E}_r$ and ${\bf H}_r$ are assumed to depend on $x$, $z$, and $t$ as $\exp(-\kappa_{1r} z+iqx-i\omega t)$,
while ${\bf E}_l$ and ${\bf H}_l$ are as $\exp(-\kappa_{1l} z+iqx-i\omega t)$.
In the $z<0$ region, ${\bf E}_r$ and ${\bf H}_r$ are proportional to $\exp(\kappa_{2r} z+iqx-i\omega t)$, while ${\bf E}_l$ and ${\bf H}_l$ are to $\exp(\kappa_{2l} z+iqx-i\omega t)$.
The imaginary wave-vector components $\kappa_{jr}$ and $\kappa_{jl}$ are defined by
\begin{eqnarray}
\kappa_{jr}=\sqrt{q^2-{k_0}^2{n_{jr}}^2},~~\kappa_{jl}=\sqrt{q^2-{k_0}^2{n_{jl}}^2},
\label{eq:kap}
\end{eqnarray}
where $k_0$ ($=\omega/c$) is the vacuum wave number.
In order to have a pure surface-wave mode, all of these components have to be positive real numbers.

The tangential components of the electric and magnetic fields have to be continuous at the interface. In addition,
the electric and magnetic fields in bi-isotropic media satisfy the relationships
\begin{eqnarray}
{\bf E}_r=i\eta_r{\bf H}_r,~~{\bf E}_l=-i\eta_l{\bf H}_l.
\label{eq:lreq}
\end{eqnarray}
Applying these conditions to the $x$ and $y$ components of the fields, we obtain
\begin{eqnarray}
&&E_{1r,y}+E_{1l,y}=E_{2r,y}+E_{2l,y},\nonumber\\
&&\frac{E_{1r,y}}{\eta_{1r}}-\frac{E_{1l,y}}{\eta_{1l}}=\frac{E_{2r,y}}{\eta_{2r}}-\frac{E_{2l,y}}{\eta_{2l}},\nonumber\\
&&E_{1r,x}+E_{1l,x}=E_{2r,x}+E_{2l,x},\nonumber\\
&&\frac{E_{1r,x}}{\eta_{1r}}-\frac{E_{1l,x}}{\eta_{1l}}=\frac{E_{2r,x}}{\eta_{2r}}-\frac{E_{2l,x}}{\eta_{2l}}.
\label{eq:ffe}
\end{eqnarray}
From Maxwell's equations, we can derive the relationship between $E_x$, $E_y$, and $H_y$:
\begin{eqnarray}
E_x(z)=-\frac{i}{k_0}\frac{\mu {H_y}^\prime+a{E_y}^\prime}{\epsilon\mu-\vert a\vert^2},
\end{eqnarray}
where the prime denotes a differentiation with respect to $z$.
Substituting this into Eq.~(\ref{eq:ffe}), we obtain a matrix equation of the form
\begin{eqnarray}
\left(\begin{array}{cccc}
1 & 1 & -1 & -1 \\ \frac{1}{\eta_{1r}} & -\frac{1}{\eta_{1l}} & -\frac{1}{\eta_{2r}} & \frac{1}{\eta_{2l}}
\\ \frac{\kappa_{1r}}{n_{1r}} & -\frac{\kappa_{1l}}{n_{1l}} & \frac{\kappa_{2r}}{n_{2r}} & -\frac{\kappa_{2l}}{n_{2l}}
\\ \frac{\kappa_{1r}}{n_{1r}\eta_{1r}} & \frac{\kappa_{1l}}{n_{1l}\eta_{1l}} & \frac{\kappa_{2r}}{n_{2r}\eta_{2r}} & \frac{\kappa_{2l}}{n_{2l}\eta_{2l}}
\end{array}\right)\left(\begin{array}{cccc}E_{1r,y}\\E_{1l,y}\\E_{2r,y}\\E_{2l,y}\end{array}\right)=0.
\label{eq:meq}
\end{eqnarray}
To have a non-trivial solution, the determinant of the $4\times 4$ coefficient matrix of this equation has to vanish.
This condition yields the desired dispersion relation for surface waves at the interface between two different
bi-isotropic media:
\begin{eqnarray}
&&\left(\eta_{1r}+\eta_{1l}\right)\left(\eta_{2r}+\eta_{2l}\right)
\left(\frac{\kappa_{1r}}{n_{1r}}+\frac{\kappa_{2r}}{n_{2r}}\right)
\left(\frac{\kappa_{1l}}{n_{1l}}+\frac{\kappa_{2l}}{n_{2l}}\right)+\nonumber\\&&
\left(\eta_{1r}-\eta_{2r}\right)\left(\eta_{1l}-\eta_{2l}\right)
\left(\frac{\kappa_{1r}}{n_{1r}}+\frac{\kappa_{1l}}{n_{1l}}\right)
\left(\frac{\kappa_{2r}}{n_{2r}}+\frac{\kappa_{2l}}{n_{2l}}\right)=0.\nonumber\\
\label{eq:disp}
\end{eqnarray}

In the special case where both of the two bi-isotropic media are Tellegen media with nonzero $\chi_1$ and $\chi_2$
but with $\gamma_1=\gamma_2=0$, we have $n_{jr}=n_{jl}=n_{j}$ and the dispersion relation can be simplified as
\begin{eqnarray}
\left(\epsilon_1\kappa_2+\epsilon_2\kappa_1\right)\left(\mu_1\kappa_2+\mu_2\kappa_1\right)
-\left(\chi_1\kappa_2+\chi_2\kappa_1\right)^2=0,\nonumber\\
\label{eq:disp2}
\end{eqnarray}
where
\begin{eqnarray}
\kappa_{j}=\sqrt{q^2-{k_0}^2{n_{j}}^2}.
\end{eqnarray}
When both of the two media are ordinary isotropic media with $\chi_1=\chi_2=\gamma_1=\gamma_2=0$,
this equation
reduces to the well-known dispersion relation
\begin{eqnarray}
\left(\epsilon_1\kappa_2+\epsilon_2\kappa_1\right)\left(\mu_1\kappa_2+\mu_2\kappa_1\right)=0.
\end{eqnarray}

For the discussion of omnidirectional total transmission through a conjugate matched pair
in the following sections, it is beneficial to introduce the effective dielectric
permittivities $\epsilon_{jr}$ and $\epsilon_{jl}$ and the effective magnetic permeabilities $\mu_{jr}$ and $\mu_{jl}$ ($j=1,2$).
Starting from Eq.~(\ref{eq:lreq}) and using the definitions
\begin{eqnarray}
&&{\bf D}_r=\epsilon_r {\bf E}_r,~~{\bf D}_l=\epsilon_l {\bf E}_l,\nonumber\\
&&{\bf B}_r=\mu_r {\bf H}_r,~~{\bf B}_l=\mu_l {\bf H}_l,
\end{eqnarray}
it is straightforward to derive the expressions
\begin{eqnarray}
&&\epsilon_{jr}=\epsilon_j+\frac{a_j}{i\eta_{jr}},~~\epsilon_{jl}=\epsilon_j-\frac{a_j}{i\eta_{jl}},\nonumber\\
&&\mu_{jr}=\mu_j+ia_j^*\eta_{jr},~~\mu_{jl}=\mu_j-ia_j^*\eta_{jl},
\label{eq:epmu}
\end{eqnarray}
using which we can rewrite Eq.~(\ref{eq:neta}) as
\begin{eqnarray}
&&n_{jr}=\sqrt{\epsilon_{jr}}\sqrt{\mu_{jr}},~~n_{jl}=\sqrt{\epsilon_{jl}}\sqrt{\mu_{jl}},\nonumber\\
&&\eta_{jr}=\frac{\sqrt{\mu_{jr}}}{\sqrt{\epsilon_{jr}}},~~\eta_{jl}=\frac{\sqrt{\mu_{jl}}}{\sqrt{\epsilon_{jl}}}.
\label{eq:impq}
\end{eqnarray}

\section{Omnidirectional excitation of surface waves and super-Klein tunneling}
\label{sec22}

\begin{table*}
\caption{\label{table1}
Summary of the conditions under which the surface waves are excited regardless of
the component of the wave vector parallel to the interface, or equivalently,
the incident angle in ATR experiments on multilayer structures.
The cases where the interface is between two Tellegen media or between two chiral media are considered.
The polarization of the excited surface-wave mode is indicated and each case is assigned a case number.
The conditions II expressed in terms of the effective dielectric permittivities and magnetic permeabilities
are equivalent to the conditions I.}
\begin{ruledtabular}
  \begin{tabular}{lllllll}
    Interface &Conditions I & &Conditions II & & Surface mode & Case no.\\
    \hline
    Tellegen/Tellegen & $n_1=-n_2$ & $\eta_{1r}=\eta_{2r}$ &$\epsilon_{1r}=-\epsilon_{2r}$ &$\mu_{1r}=-\mu_{2r}$ & $(1,0,1,0)$ &I\\
     & $n_1=-n_2$ & $\eta_{1l}=\eta_{2l}$ &$\epsilon_{1l}=-\epsilon_{2l}$ &$\mu_{1l}=-\mu_{2l}$ & $(0,1,0,1)$ &II\\
     & $n_1=n_2$ & $\eta_{1r}=-\eta_{2l}$ &$\epsilon_{1r}=-\epsilon_{2l}$ &$\mu_{1r}=-\mu_{2l}$ &$(1,0,0,1)$ &III\\
     & $n_1=n_2$ & $\eta_{1l}=-\eta_{2r}$ &$\epsilon_{1l}=-\epsilon_{2r}$ &$\mu_{1l}=-\mu_{2r}$ &$(0,1,1,0)$ &IV\\
     \hline
    Chiral/Chiral & $\eta_1=-\eta_2$ & $n_{1r}=n_{2l}$ &$\epsilon_{1r}=-\epsilon_{2l}$ &$\mu_{1r}=-\mu_{2l}$ &$(1,0,0,1)$ &V\\
     & $\eta_1=-\eta_2$ & $n_{1l}=n_{2r}$ &$\epsilon_{1l}=-\epsilon_{2r}$ &$\mu_{1l}=-\mu_{2r}$ &$(0,1,1,0)$ &VI\\
     & $\eta_1=\eta_2$ & $n_{1r}=-n_{2r}$ &$\epsilon_{1r}=-\epsilon_{2r}$ &$\mu_{1r}=-\mu_{2r}$ & $(1,0,1,0)$ &VII\\
     & $\eta_1=\eta_2$ & $n_{1l}=-n_{2l}$ &$\epsilon_{1l}=-\epsilon_{2l}$ &$\mu_{1l}=-\mu_{2l}$ & $(0,1,0,1)$ &VIII\\
  \end{tabular}

  \end{ruledtabular}
\end{table*}

In this section, we derive some interesting consequences of the analytical dispersion relation.
In particular, we derive the conditions under which the surface waves are excited regardless of the incident angle
in ATR experiments on multilayer structures
similar to the Kretschmann or Otto configuration.

When both of the two bi-isotropic media are Tellegen media, the dispersion relation [Eq.~(\ref{eq:disp})] is reduced to
\begin{eqnarray}
&&\left(\eta_{1r}+\eta_{1l}\right)\left(\eta_{2r}+\eta_{2l}\right)
\left(\frac{\kappa_{1}}{n_{1}}+\frac{\kappa_{2}}{n_{2}}\right)^2\nonumber\\&&
~+4\left(\eta_{1r}-\eta_{2r}\right)\left(\eta_{1l}-\eta_{2l}\right)
\frac{\kappa_{1}}{n_{1}}\frac{\kappa_{2}}{n_{2}}=0.
\end{eqnarray}
If the effective refractive indices of the two media are anti-matched (that is, $n_1=-n_2$ and $\kappa_1=\kappa_2$), then this equation becomes
\begin{eqnarray}
\left(\eta_{1r}-\eta_{2r}\right)\left(\eta_{1l}-\eta_{2l}\right)\left(\frac{{\kappa_1}}{{n_1}}\right)^2=0.
\end{eqnarray}
We notice that it is satisfied for any value of $\kappa_1$ if
\begin{eqnarray}
\eta_{1r}=\eta_{2r}~ \text{ or }~ \eta_{1l}=\eta_{2l}.
\label{eq:cq1}
\end{eqnarray}
Since the dependence on the incident angle in ATR experiments occurs only through $\kappa_j$, we conclude that
surface waves will be excited regardless of the incident angle if any of the above conditions is satisfied, as long as the incident angle
satisfies the condition that $\kappa_j$ is real. This makes surface waves be excited at all or a continuous range of incident angles,
in sharp contrast to the usual case where surface waves are excited at a specific incident angle.

On the other hand,
if the effective refractive indices of the two Tellegen media are matched (that is, $n_1=n_2$ and $\kappa_1=\kappa_2$),
the dispersion relation is reduced to
\begin{eqnarray}
\left(\eta_{1r}+\eta_{2l}\right)\left(\eta_{1l}+\eta_{2r}\right)\left(\frac{{\kappa_1}}{{n_1}}\right)^2=0,
\end{eqnarray}
which is satisfied for any (real) value of $\kappa_1$ if
\begin{eqnarray}
\eta_{1r}=-\eta_{2l}~ \text{ or }~ \eta_{1l}=-\eta_{2r}.
\label{eq:cq2}
\end{eqnarray}

Similarly, when both media are chiral media such that $\chi_1=\chi_2=0$ and $\eta_{jr}=\eta_{jl}\equiv\eta_j$,
the dispersion relation [Eq.~(\ref{eq:disp})] is reduced to
\begin{eqnarray}
&&4\eta_1\eta_2\left(\frac{\kappa_{1r}}{n_{1r}}+\frac{\kappa_{2r}}{n_{2r}}\right)
\left(\frac{\kappa_{1l}}{n_{1l}}+\frac{\kappa_{2l}}{n_{2l}}\right)\nonumber\\
&&~+\left(\eta_1-\eta_2\right)^2\left(\frac{\kappa_{1r}}{n_{1r}}+\frac{\kappa_{1l}}{n_{1l}}\right)
\left(\frac{\kappa_{2r}}{n_{2r}}+\frac{\kappa_{2l}}{n_{2l}}\right)=0.
\end{eqnarray}
If the impedances of the two media are anti-matched (that is, $\eta_1=-\eta_2$), then this equation becomes
\begin{eqnarray}
{\eta_1}^2\left(\frac{\kappa_{1r}}{n_{1r}}-\frac{\kappa_{2l}}{n_{2l}}\right)
\left(\frac{\kappa_{1l}}{n_{1l}}-\frac{\kappa_{2r}}{n_{2r}}\right)=0,
\end{eqnarray}
which is satisfied identically if
\begin{eqnarray}
n_{1r}=n_{2l}~ \text{ or }~ n_{1l}=n_{2r}.
\label{eq:cq3}
\end{eqnarray}

On the other hand, if the impedances of the two media are matched (that is, $\eta_1=\eta_2$), the dispersion relation is reduced to
\begin{eqnarray}
{\eta_1}^2\left(\frac{\kappa_{1r}}{n_{1r}}+\frac{\kappa_{2r}}{n_{2r}}\right)
\left(\frac{\kappa_{1l}}{n_{1l}}+\frac{\kappa_{2l}}{n_{2l}}\right)=0,
\end{eqnarray}
which is satisfied identically if
\begin{eqnarray}
n_{1r}=-n_{2r}~ \text{ or }~ n_{1l}=-n_{2l}.
\label{eq:cq4}
\end{eqnarray}

In Table~\ref{table1} we give a summary of the conditions under which the surface waves are excited regardless of
the component of the wave vector parallel to the interface, or equivalently, the incident angle
in ATR experiments on multilayer structures.
We also indicate the polarization of the excited surface-wave mode.
For instance, when the conditions $n_1=-n_2$ and $\eta_{1r}=\eta_{2r}$ are satisfied for two Tellegen media,
we can easily verify from Eq.~(\ref{eq:meq}) that the corresponding surface-wave mode satisfies
\begin{eqnarray}
\left(E_{1r,y},E_{1l,y},E_{2r,y},E_{2l,y}\right)\propto (1,0,1,0).
\end{eqnarray}
This implies that the surface wave is RCP in both media 1 and 2.
The polarizations in other cases are obtained similarly.
In the situation where the surface wave is excited by a wave incident from the side of the medium 1, the circular
polarization of the wave reflected from the interface has to match with that of the excited surface-wave mode. 
An example of this phenomenon will be discussed in Sec.~\ref{ss_a}.

The conditions II listed in Table~\ref{table1} are expressed in terms of the effective dielectric permittivities and magnetic permeabilities
and have been derived from the conditions I using Eq.~(\ref{eq:impq}). For instance, from $n_1=-n_2$ and $\eta_{1r}=\eta_{2r}$,
we have
\begin{eqnarray}
\sqrt{\epsilon_{1r}}\sqrt{\mu_{1r}}=-\sqrt{\epsilon_{2r}}\sqrt{\mu_{2r}},~~
\frac{\sqrt{\mu_{1r}}}{\sqrt{\epsilon_{1r}}}=\frac{\sqrt{\mu_{2r}}}{\sqrt{\epsilon_{2r}}}.
\end{eqnarray}
From a simple manipulation of these equations, we obtain
\begin{eqnarray}
\epsilon_{1r}=-\epsilon_{2r},~~\mu_{1r}=-\mu_{2r}.
\end{eqnarray}

We notice that the conditions II are closely related to those for the super-Klein tunneling or the omnidirectional
total transmission to occur. The generalized definition of a conjugate matched pair is that the two layers have the same thicknesses
and satisfy
\begin{eqnarray}
\frac{\epsilon_{1r}}{\epsilon_{2r}}=\frac{\mu_{1r}}{\mu_{2r}}=-1~~{\rm and}~~\frac{\epsilon_{1l}}{\epsilon_{2l}}=\frac{\mu_{1l}}{\mu_{2l}}=-1
\label{eq:cmp1}
\end{eqnarray}
or
\begin{eqnarray}
\frac{\epsilon_{1r}}{\epsilon_{2l}}=\frac{\mu_{1r}}{\mu_{2l}}=-1~~{\rm and}~~\frac{\epsilon_{1l}}{\epsilon_{2r}}=\frac{\mu_{1l}}{\mu_{2r}}=-1.
\label{eq:cmp2}
\end{eqnarray}
In order for the transmittance to be equal to one for any incident angle and polarization, an additional condition that the incident and transmitted
regions are consisted of the same media has to be satisfied.

In the special case of a bilayer made of two different chiral media, RCP and LCP waves become completely decoupled,
if the incident and transmitted regions and the two chiral media have the same impedance. Then the super-Klein tunneling is achieved
separately for RCP and LCP waves when
\begin{eqnarray}
&&\frac{\epsilon_{1r}}{\epsilon_{2r}}=\frac{\mu_{1r}}{\mu_{2r}}=-1, ~~{\rm for~ RCP},\nonumber\\
&&\frac{\epsilon_{1l}}{\epsilon_{2l}}=\frac{\mu_{1l}}{\mu_{2l}}=-1, ~~{\rm for~ LCP}.
\label{eq:cmp3}
\end{eqnarray}

In dispersive media, the parameters $\epsilon$, $\mu$, $\chi$, and $\gamma$ depend on the wave frequency.
Therefore each of the matching (or anti-matching) conditions listed in Table~\ref{table1} can be satisfied only for a specific frequency in dispersive cases.
That the dispersion relation is satisfied for any value of $q$ at a certain frequency appears to have a close resemblance to the {\it flat band} phenomenon attracting
much current interest of researchers \cite{flach}.

\section{Invariant imbedding method}
\label{sec3}

\begin{figure}
\centering\includegraphics[width=8.5cm]{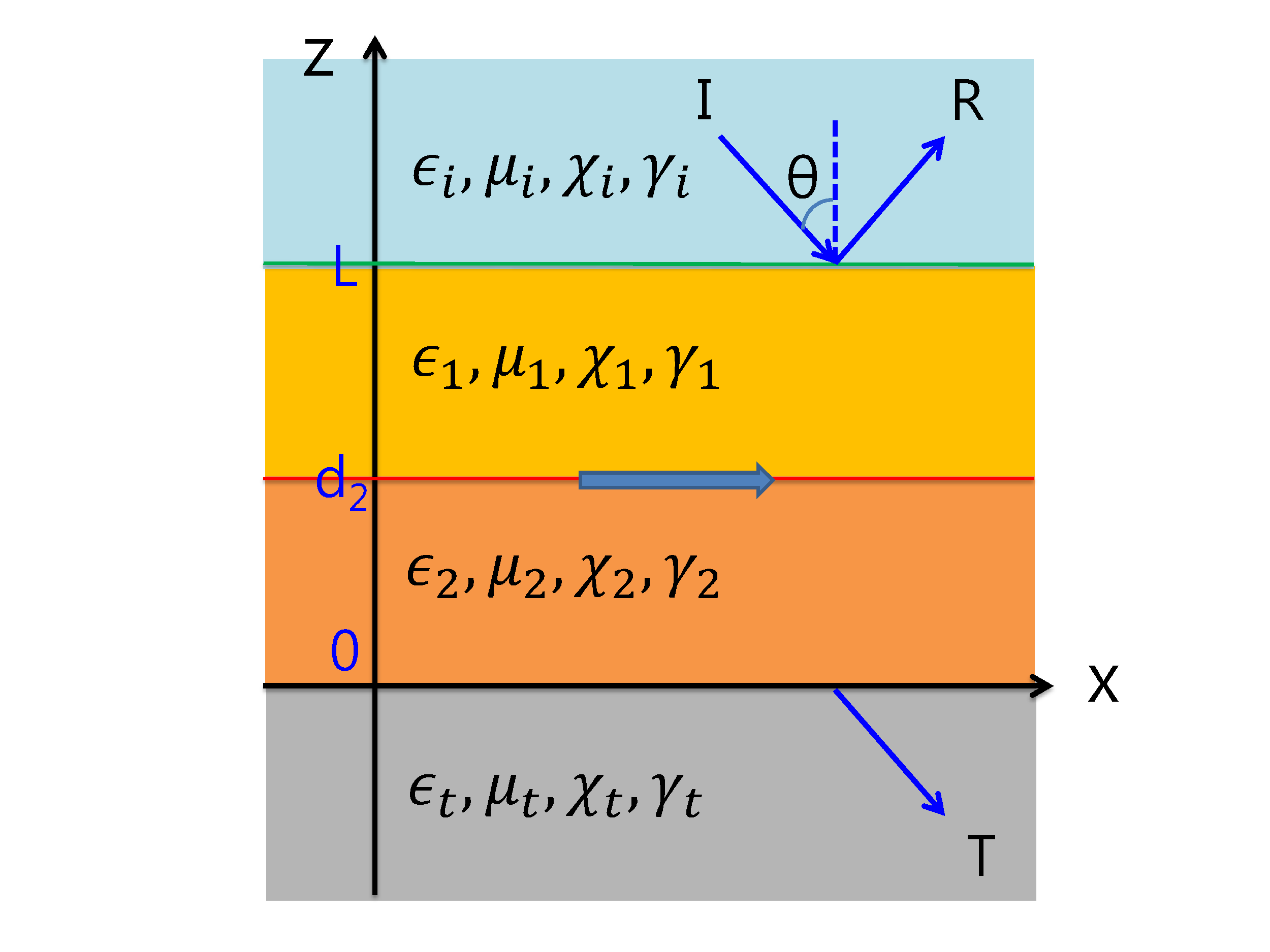}
\caption{Sketch of the configuration considered in Sec.~\ref{sec3}. A plane wave is incident at an angle $\theta$
from a bi-isotropic medium with the parameters $\epsilon_i$, $\mu_i$, $\chi_i$, and $\gamma_i$ onto a bilayer system
consisting of two different kinds of bi-isotropic media with the parameters $\epsilon_1$, $\mu_1$, $\chi_1$, and $\gamma_1$ and
$\epsilon_2$, $\mu_2$, $\chi_2$, and $\gamma_2$ respectively, and then is transmitted to the bi-isotropic substrate
with $\epsilon_t$, $\mu_t$, $\chi_t$, and $\gamma_t$. The wave is evanescent in the regions $d_2<z<L$ and $0<z<d_2$. The
incident wave tunnels through the top layer, excites the surface wave on the interface at $z=d_2$, and then tunnels through the bottom layer
to reach the substrate.}
\label{fig2}
\end{figure}

A popular method to excite surface waves experimentally is to perform ATR experiments on multilayer structures
similar to the Kretschmann or Otto configuration.
In this paper, we consider a configuration sketched in Fig.~\ref{fig2}. A plane wave is assumed to be incident at an angle $\theta$
from a uniform bi-isotropic medium with the parameters $\epsilon_i$, $\mu_i$, and $a_i$ ($=\chi_i+i\gamma_i$) onto a bilayer system
consisting of two different kinds of bi-isotropic media with the parameters $\epsilon_1$, $\mu_1$, and $a_1$ ($=\chi_1+i\gamma_1$) and
$\epsilon_2$, $\mu_2$, and $a_2$ ($=\chi_2+i\gamma_2$) respectively, and then is transmitted to the bi-isotropic substrate with $\epsilon_t$, $\mu_t$, and $a_t$
($=\chi_t+i\gamma_t$).
The layers 1 and 2 are assumed to have thicknesses $d_1$ and $d_2$ such that $d_1+d_2=L$.
The wave is evanescent in the regions $d_2<z<L$ and $0<z<d_2$. The
incident wave tunnels through the top layer, excites the surface wave on the interface at $z=d_2$, and then tunnels through the bottom layer
to reach the substrate. Some examples of the spatial field distribution in the region $0<z<L$ will be shown in Figs.~\ref{fig12}, \ref{ffgg}, and
\ref{fig32}.

We will test the predictions of the dispersion relation derived in Secs.~\ref{sec2} and \ref{sec22} by solving Maxwell's equations directly
in the configuration of Fig.~\ref{fig2}
using the IIM. In this section, we present a generalization of the IIM developed previously in Ref.~\cite{15} to the case where
the incident and transmitted regions as well as the inhomogeneously stratified medium in between are general bi-isotropic media.

As we have mentioned before, the field components $E_y$ and $H_y$ can be decomposed as
\begin{eqnarray}
&&E_y=E_{r,y}+E_{l,y},\nonumber\\
&&H_y=H_{r,y}+H_{l,y}=\frac{E_{r,y}}{i\eta_r}-\frac{E_{l,y}}{i\eta_l}.
\end{eqnarray}
We rewrite this as a matrix equation
\begin{eqnarray}
\psi=N\phi,
\label{eq:cb}
\end{eqnarray}
where
\begin{eqnarray}
 &&\psi=\pmatrix{E_y \cr H_y},~~ \phi=\pmatrix{E_{r,y} \cr E_{l,y}},\nonumber\\&& N=\pmatrix{1 & 1 \cr \frac{1}{i\eta_r} & -\frac{1}{i\eta_l}}.
\end{eqnarray}

In Ref.~\cite{15}, we have derived the matrix wave equation for $\psi$ in general bi-isotropic media of the form
\begin{eqnarray}
\left(\mathcal{E}^{-1}\psi^\prime\right)^\prime+\mathcal{D}\psi=0,
\label{eq:wwo}
\end{eqnarray}
where
\begin{eqnarray}
 &&\mathcal{D}={k_0}^2\mathcal{M}-q^2\mathcal{E}^{-1},~~\mathcal{E}=\pmatrix{\mu & -a^* \cr -a & \epsilon},\nonumber\\
 &&\mathcal{M}=\pmatrix{\epsilon & a \cr a^* & \mu}.
\end{eqnarray}
By substituting Eq.~(\ref{eq:cb}) into Eq.~(\ref{eq:wwo}), we change the basis of the wave function to $\phi$:
\begin{eqnarray}
\left[\mathcal{E}^{-1}\left(N\phi\right)^\prime\right]^\prime+\mathcal{D}N\phi=0.
\label{eq:we3}
\end{eqnarray}
It is necessary to generalize
this equation by replacing the vector wave function
$\phi$ by the $2\times 2$ matrix wave function $\Phi$, the $j$th
column vector $(\Phi_{1j},\Phi_{2j})^{\rm T}$ of which represents
the wave function when the incident wave consists only of the $j$th
wave ($j=1,2$). We note that here the index $j=1$ ($j=2$) corresponds to the case where RCP (LCP) waves are incident.
We are mainly interested in calculating the $2\times 2$ reflection and
transmission coefficient matrices $r=r(L)$ and $t=t(L)$, which we consider as functions of $L$.
In our notation,
$r_{21}$ is the reflection coefficient
when the incident wave is RCP and the reflected wave is LCP.
Similarly, $r_{12}$ is the reflection coefficient
when the incident wave is LCP and the reflected wave is RCP.
Similar definitions are applied to the transmission coefficients.

We also need to generalize the vector wave function $\psi$ to the $2\times 2$ matrix wave function $\Psi$ ($= N\Phi$)
in a similar manner. The $j$th column vector $(\Psi_{1j},\Psi_{2j})^{\rm T}$ of $\Psi$
represents the wave function when the incident wave consists only of the $j$th wave. $\Psi_{11}$ ($\Psi_{12}$) represents the field $E_y$
and $\Psi_{21}$ ($\Psi_{22}$) represents the field $H_y$ in the inhomogeneous region when the incident wave is RCP (LCP), respectively.
The field $E_y$ associated with the incident wave is assumed to have a unit amplitude.

We follow the standard procedure to derive the invariant imbedding equations.
We introduce $2\times 2$ matrix functions
\begin{eqnarray}
U_1(z;L)=N\Phi,~~U_2(z;L)={\cal E}^{-1}\left(N\Phi\right)^\prime,
\end{eqnarray}
which we consider as functions of both $z$ and $L$. Then the wave equation is transformed to
\begin{eqnarray}
 \pmatrix{U_1 \cr U_2}^\prime=A\pmatrix{U_1 \cr U_2},~~A=\pmatrix{ O & \mathcal{E} \cr -\mathcal{D} & O},
\end{eqnarray}
where $A$ is a $4\times 4$ matrix and $O$ is the $2\times 2$ null matrix.

The wave functions in the incident and transmitted regions are expressed in
terms of $r$ and $t$:
\begin{eqnarray}
U_1(z;L)=\left\{ \begin{array}{ll}N_i e^{iP_i(L-z)}I +N_ie^{iP_i(z-L)}~r,
&~z>L\\
N_t e^{-iP_t z}~t, &~z<0 \end{array} \right., \nonumber\\
\label{eq:psi}
\end{eqnarray}
where $I$ is the $2\times 2$ identity matrix
and the matrices $N_i$ and $N_t$ are the values of $N$ in the incident and transmitted regions, respectively.
The diagonal matrices $P_i$ and $P_t$ are defined by
\begin{eqnarray}
 P_i=\pmatrix{ p_{ir} & 0 \cr 0 & p_{il}},~~P_t=\pmatrix{ p_{tr} & 0 \cr 0 & p_{tl}},
\label{eq:as1}
\end{eqnarray}
where the {\it negative} $z$ components of the wave vector
in the incident region $p_{ir}$ and $p_{il}$ and those in the transmitted region $p_{tr}$ and $p_{tl}$ are given by
\begin{eqnarray}
&&p_{ir} ={\rm sgn}\left(n_{ir}\right)\sqrt{{k_0}^2{n_{ir}}^2-q^2},\nonumber\\
&&p_{il} ={\rm sgn}\left(n_{il}\right)\sqrt{{k_0}^2{n_{il}}^2-q^2},\nonumber\\
&&p_{tr} =\left\{ \begin{array}{ll} {\rm sgn}\left(n_{tr}\right)\sqrt{{k_0}^2{n_{tr}}^2-q^2},& {\mbox {if}}~~ {k_0}^2{n_{tr}}^2\ge q^2\\
i\sqrt{q^2-{k_0}^2{n_{tr}}^2},& {\mbox {if}}~~{k_0}^2{n_{tr}}^2< q^2\end{array} \right.,\nonumber\\
&&p_{tl} =\left\{ \begin{array}{ll} {\rm sgn}\left(n_{tl}\right)\sqrt{{k_0}^2{n_{tl}}^2-q^2},& {\mbox {if}}~~ {k_0}^2{n_{tl}}^2\ge q^2\\
i\sqrt{q^2-{k_0}^2{n_{tl}}^2},& {\mbox {if}}~~{k_0}^2{n_{tl}}^2< q^2\end{array} \right..\nonumber\\
\label{eq:as2}
\end{eqnarray}
The effective refractive indices $n_{ir}$, $n_{il}$, $n_{tr}$, and $n_{tl}$ are defined similarly as in Eq.~(\ref{eq:neta}).
In the simpler case where the incident and transmitted regions are ordinary dielectric media, we can simplify Eqs.~(\ref{eq:as1}) and (\ref{eq:as2})
as
\begin{eqnarray}
&&P_i=p_iI,~~P_t=p_tI,\nonumber\\
&&p_i={\rm sgn}\left(n_{i}\right)\sqrt{{k_0}^2{n_{i}}^2-q^2},\nonumber\\
&&p_t=\left\{ \begin{array}{ll} {\rm sgn}\left(n_{t}\right)\sqrt{{k_0}^2{n_{t}}^2-q^2},& {\mbox {if}}~~ {k_0}^2{n_{t}}^2\ge q^2\\
i\sqrt{q^2-{k_0}^2{n_{t}}^2},& {\mbox {if}}~~{k_0}^2{n_{t}}^2< q^2\end{array} \right..
\end{eqnarray}

At the boundaries of the inhomogeneous medium, we have
\begin{eqnarray}
&&U_1(0;L)=N_tt,~~~U_1(L;L)=N_i\left(r+I\right),\nonumber\\
&&U_2(0;L)=-i{{\mathcal{E}}_t}^{-1}N_t P_t t=-i{{\mathcal{E}}_t}^{-1}N_t P_t {N_t}^{-1}U_1(0;L),\nonumber\\
&&U_2(L;L)=i{{\mathcal{E}}_i}^{-1}N_i P_i\left(r-I\right)\nonumber\\
&&~~~~~~~~~~~=i{{\mathcal{E}}_i}^{-1}N_i P_i{N_i}^{-1}U_1(L;L)-2i{{\mathcal{E}}_i}^{-1}N_i P_i,
\label{eq:bcz}
\end{eqnarray}
where ${\mathcal E}_i$ and ${\mathcal E}_t$ are the values of $\mathcal E$ in the incident and transmitted regions respectively.
From Eq.~(\ref{eq:bcz}), we obtain
\begin{eqnarray}
g{\hat S}+h{\hat R}=v,
\end{eqnarray}
where
\begin{eqnarray}
 &&{\hat S}=\pmatrix{U_1(0;L) \cr U_2(0;L)}, ~~~{\hat R}=\pmatrix{U_1(L;L) \cr U_2(L;L)},\nonumber\\
 &&g=\pmatrix{i{{\mathcal{E}}_t}^{-1}N_t P_t {N_t}^{-1} & I \cr O & O},\nonumber\\
 &&h=\pmatrix{ O & O \cr i{{\mathcal{E}}_i}^{-1}N_i P_i{N_i}^{-1} & -I},\nonumber\\
 &&v=\pmatrix{ O \cr 2i{{\mathcal{E}}_i}^{-1}N_i P_i}.
\end{eqnarray}
We define $4 \times 4$ matrices $\mathcal S$ and $\mathcal R$ by
\begin{eqnarray}
 {\hat S}={\mathcal S}v=\pmatrix{S_{11} & S_{12} \cr S_{21} & S_{22}}v,~~{\hat R}={\mathcal R}v=\pmatrix{R_{11} & R_{12} \cr R_{21} & R_{22}}v,
\nonumber\\
\end{eqnarray}
where $S_{ij}$ and $R_{ij}$ ($i,j=1,2$) are $2\times 2$ matrices.
The invariant imbedding equations satisfied by ${\mathcal R}$ and ${\mathcal S}$ have been derived in Ref.~\cite{15}:
\begin{eqnarray}
&&\frac{d{\mathcal R}}{dl}=A(l){\mathcal R}(l)-{\mathcal R}(l)h A(l){\mathcal R}(l),\nonumber\\
&&\frac{d{\mathcal S}}{dl}=-{\mathcal S}(l)h A(l){\mathcal R}(l),
\label{eq:iea}
\end{eqnarray}
where $l$ is the thickness of the inhomogeneous layer in the $z$ direction.
The initial conditions for these matrices are given by
\begin{eqnarray}
{\mathcal R}(0)={\mathcal S}(0)=\left(g+h\right)^{-1}.
\label{eq:icaa}
\end{eqnarray}
From the definitions of $\hat R$, $\hat S$, $\mathcal R$, and $\mathcal S$, we obtain
\begin{eqnarray}
&&R_{12}\left(2i{{\mathcal{E}}_i}^{-1}N_i P_i\right)=N_i\left(r+I\right),\nonumber\\
&&R_{22}\left(2i{{\mathcal{E}}_i}^{-1}N_i P_i\right)=i{{\mathcal{E}}_i}^{-1}N_i P_i\left(r-I\right),\nonumber\\
&&S_{12}\left(2i{{\mathcal{E}}_i}^{-1}N_i P_i\right)=N_t t,\nonumber\\
&&S_{22}\left(2i{{\mathcal{E}}_i}^{-1}N_i P_i\right)=-i{{\mathcal{E}}_t}^{-1}N_t P_t t.
\label{eq:ieb}
\end{eqnarray}
The expression for $S_{22}$ is given for reference, though it is not necessary for the derivation of the invariant imbedding equations.

The invariant imbedding equations satisfied by $r$ and $t$ follow from Eqs.~(\ref{eq:iea}) and (\ref{eq:ieb})
and take the forms
\begin{eqnarray}
&&\frac{dr}{dl}=i\left({N_i}^{-1}\mathcal{E}{\mathcal{E}_i}^{-1}N_iP_ir+r{N_i}^{-1}\mathcal{E}{\mathcal{E}_i}^{-1}N_iP_i\right) \nonumber\\
&&-\frac{i}{2}\left(r+I\right)\left({N_i}^{-1}\mathcal{E}{\mathcal{E}_i}^{-1}N_iP_i
-{P_i}^{-1}{N_i}^{-1}{\mathcal{E}_i}{\mathcal D}N_i\right)\left(r+I\right),\nonumber\\
&&\frac{dt}{dl}=it{N_i}^{-1}\mathcal{E}{\mathcal{E}_i}^{-1}N_iP_i \nonumber\\
&&-\frac{i}{2}t\left({N_i}^{-1}\mathcal{E}{\mathcal{E}_i}^{-1}N_iP_i
-{P_i}^{-1}{N_i}^{-1}{\mathcal{E}_i}{\mathcal D}N_i\right)\left(r+I\right).
\end{eqnarray}
The initial conditions for $r$ and $t$ are obtained from Eq.~(\ref{eq:icaa}):
\begin{widetext}
\begin{eqnarray}
&&r(0)=2{N_i}^{-1}\left({\mathcal{E}_i}^{-1}N_iP_i{N_i}^{-1}+{\mathcal{E}_t}^{-1}N_tP_t{N_t}^{-1}\right)^{-1}{\mathcal{E}_i}^{-1}N_iP_i-I,\nonumber\\
&&t(0)=2{N_t}^{-1}\left({\mathcal{E}_i}^{-1}N_iP_i{N_i}^{-1}+{\mathcal{E}_t}^{-1}N_tP_t{N_t}^{-1}\right)^{-1}{\mathcal{E}_i}^{-1}N_iP_i.
\end{eqnarray}
\end{widetext}

The invariant imbedding method can also be used in calculating the wave function $\Psi(z;L)$ inside the inhomogeneous medium.
It turns out that the equation satisfied by $\Psi(z;L)$ is very similar to that for $t$ and takes the form
\begin{eqnarray}
&&\frac{\partial}{\partial l}\Psi(z;l)=i\Psi{N_i}^{-1}\mathcal{E}{\mathcal{E}_i}^{-1}N_iP_i \nonumber\\
&&~~-\frac{i}{2}\Psi\left({N_i}^{-1}\mathcal{E}{\mathcal{E}_i}^{-1}N_iP_i
-{P_i}^{-1}{N_i}^{-1}{\mathcal{E}_i}{\mathcal D}N_i\right)\left(r+I\right).\nonumber\\
\end{eqnarray}
This equation is integrated from $l=z$ to $l=L$ using the initial condition
\begin{eqnarray}
\Psi(z;z)=N_i\left[r(z)+I\right].
\end{eqnarray}

The reflectance (transmittance) is defined by the ratio of the energy flux of the reflected (transmitted) wave to that of the incident wave.
The energy flux of a wave is given by the $z$ component of the Poynting vector
\begin{eqnarray}
{\rm Re}\left(S_z\right)&=&\frac{c}{4\pi}{\rm Re}\left[\left({\bf E}\times {\bf H}^*\right)_z\right]\nonumber\\
&=&\frac{c}{4\pi}{\rm Re}\left(E_x H_y^*-E_y H_x^*\right).
\end{eqnarray}
Starting from this, we obtain the expressions for the components of the $2\times 2$ reflectance and transmittance matrices of the form
\begin{eqnarray}
&&R_{11}=\vert r_{11}\vert^2,~~R_{21}=\frac{p_{il}n_{ir}}{p_{ir}n_{il}}\vert r_{21}\vert^2,\nonumber\\
&&R_{12}=\frac{p_{ir}n_{il}}{p_{il}n_{ir}}\vert r_{12}\vert^2,~~R_{22}=\vert r_{22}\vert^2,\nonumber\\
&&T_{11}=\frac{p_{tr}n_{ir}\mu_i\sqrt{\epsilon_t\mu_t-{\chi_t}^2}}{p_{ir}n_{tr}\mu_t\sqrt{\epsilon_i\mu_i-{\chi_i}^2}}
\vert t_{11}\vert ^2,\nonumber\\
&&T_{21}=\frac{p_{tl}n_{ir}\mu_i\sqrt{\epsilon_t\mu_t-{\chi_t}^2}}{p_{ir}n_{tl}\mu_t\sqrt{\epsilon_i\mu_i-{\chi_i}^2}}
\vert t_{21}\vert ^2,\nonumber\\
&&T_{12}=\frac{p_{tr}n_{il}\mu_i\sqrt{\epsilon_t\mu_t-{\chi_t}^2}}{p_{il}n_{tr}\mu_t\sqrt{\epsilon_i\mu_i-{\chi_i}^2}}
\vert t_{12}\vert ^2,\nonumber\\
&&T_{22}=\frac{p_{tl}n_{il}\mu_i\sqrt{\epsilon_t\mu_t-{\chi_t}^2}}{p_{il}n_{tl}\mu_t\sqrt{\epsilon_i\mu_i-{\chi_i}^2}}
\vert t_{22}\vert ^2.
\label{eq:as3}
\end{eqnarray}
In the case where $p_{tr}$ ($p_{tl}$) is imaginary, we have to set $T_{11}=T_{12}=0$ ($T_{21}=T_{22}=0$),
since the transmitted wave is evanescent.
On the other hand, if both $\epsilon_t$ and $\mu_t$ are negative, while $\epsilon_t\mu_t>{\chi_t}^2$,
we need to replace $\sqrt{\epsilon_t\mu_t-{\chi_t}^2}$ by $-\sqrt{\epsilon_t\mu_t-{\chi_t}^2}$ in the above expressions
for the transmittances.
In the simpler case where the incident and transmitted regions are ordinary dielectric media, we can simplify Eq.~(\ref{eq:as3})
as
\begin{eqnarray}
R_{ij}=\vert r_{ij}\vert^2,~~T_{ij}=\frac{p_t\mu_i}{p_i\mu_t}\vert t_{ij}\vert^2.
\end{eqnarray}
We stress again that the numerical indices 1 and 2 refer to the RCP and LCP waves, respectively,
and the first index $i$ of the matrix $R_{ij}$ ($T_{ij}$)
is used for the reflected (transmitted) wave, while
the second index $j$ is used for the incident wave. For example, $T_{12}$ represents the transmittance when the incident wave is LCP
and the transmitted wave is RCP.

In the absence of dissipation, the law of energy conservation requires that
\begin{eqnarray}
&&R_{11}+R_{21}+T_{11}+T_{21}=1,\nonumber\\
&&R_{22}+R_{12}+T_{22}+T_{12}=1.
\end{eqnarray}
If there is dissipation, then the absorptance is defined by
\begin{eqnarray}
&&A_1=1-R_{11}-R_{21}-T_{11}-T_{21},\nonumber\\
&&A_2=1-R_{22}-R_{12}-T_{22}-T_{12},
\end{eqnarray}
where $A_1$ ($A_2$) is the absorptance when the incident wave is RCP (LCP).

\section{Numerical results}
\label{sec4}

\begin{figure}
\centering\includegraphics[width=8.5cm]{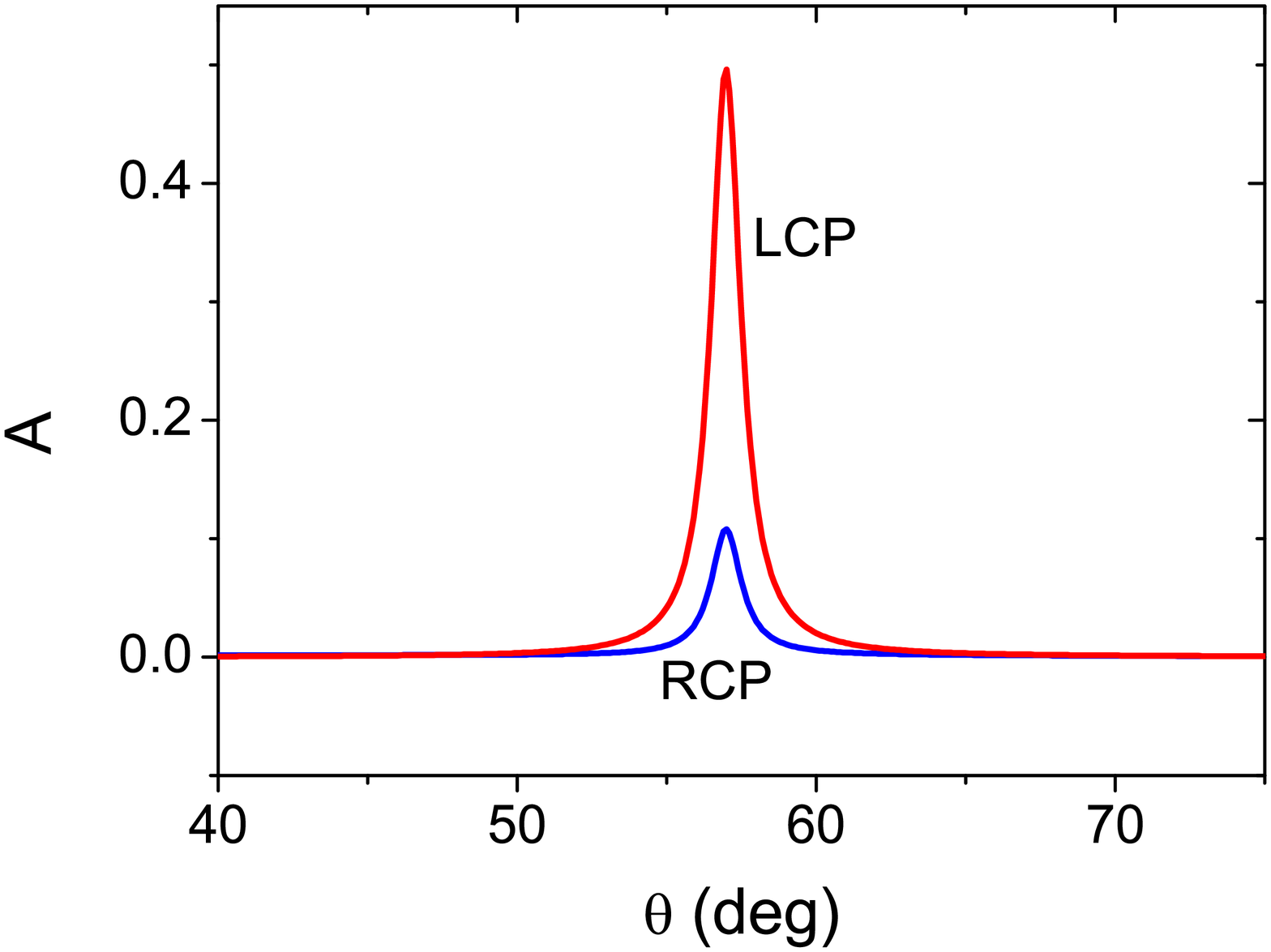}
\caption{Absorptances $A_1$ and $A_2$ for RCP and LCP waves of frequency $\omega$ incident on a bilayer system consisting of
a Tellegen medium with $\epsilon_1=2.25+0.01i$, $\mu_1=1$, $\chi_1=4$, and $\gamma_1=0$ and an ordinary dielectric
with $\epsilon_2=1.5$, $\mu_2=1$, and $\chi_2=\gamma_2=0$ in the configuration shown in Fig.~\ref{fig2} plotted versus incident angle.
Waves are incident from a dielectric prism with the parameters $\epsilon_i=9$, $\mu_i=1$, and $a_i=0$. The substrate is assumed to be made of the
same material as the prism
(that is, $\epsilon_t=9$, $\mu_t=1$, and $a_t=0$). The layers 1 and 2 have the thicknesses $d_1=\Lambda$ and $d_2=4\Lambda$,
where $\Lambda$ satisfies $\omega\Lambda/c=0.2\pi$.}
\label{fig11}
\end{figure}

\begin{figure}
\centering\includegraphics[width=8.5cm]{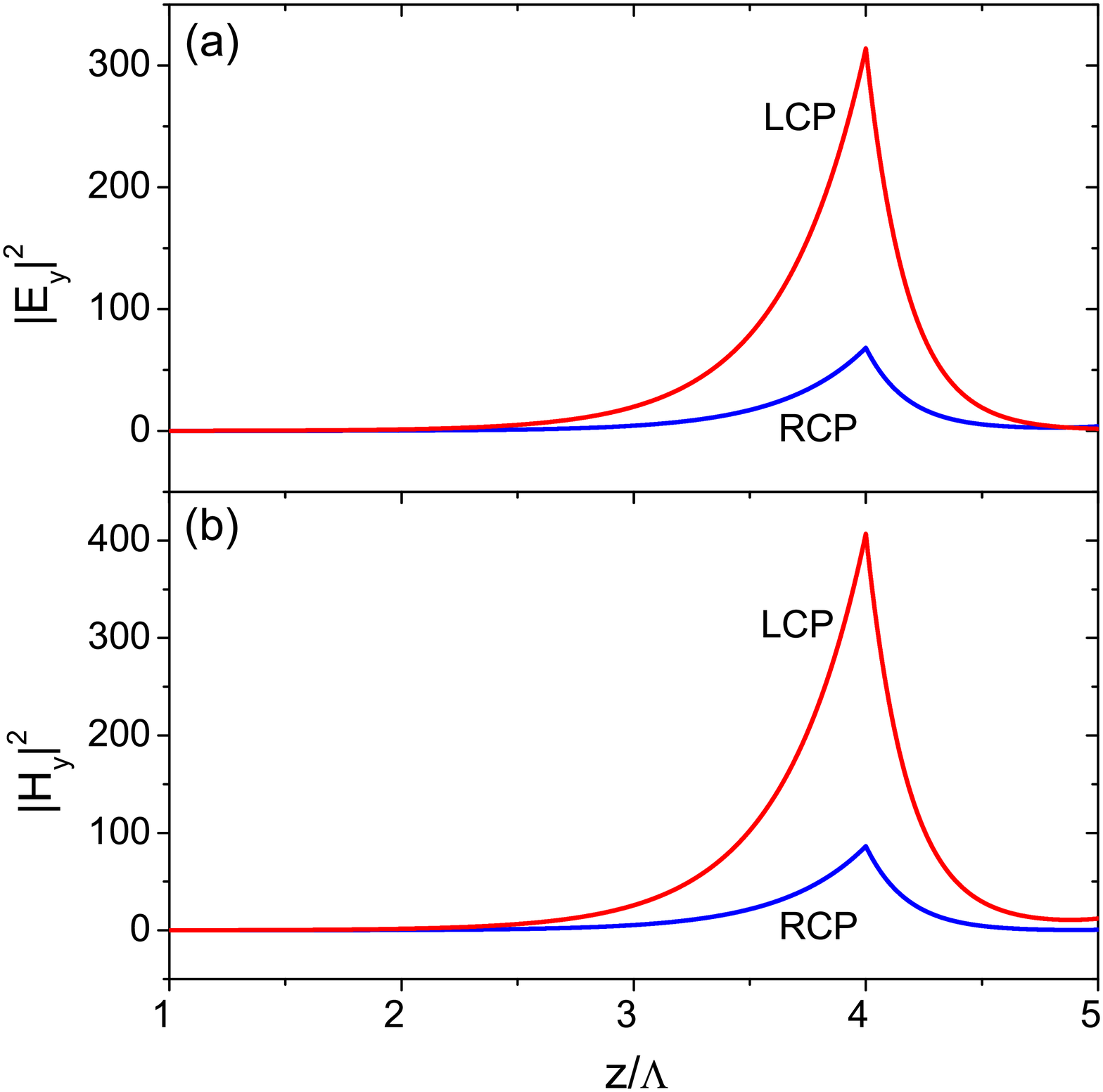}
\caption{Spatial distributions of the intensities of the $y$ components
of the electric and magnetic fields, $\vert E_y\vert^2$ and $\vert H_y\vert^2$, when RCP and LCP waves
are incident on the same bilayer system considered in Fig.~\ref{fig11}.
The $E_y$ field associated with the incident wave is assumed to have a unit amplitude.
The incident angle is chosen to be $\theta=57^\circ$.
The interface is located at $z=4\Lambda$ and waves are incident from the region where $z>5\Lambda$.}
\label{fig12}
\end{figure}

The IIM developed in Sec.~\ref{sec3} allows us to solve any wave propagation problem in the situation where
the medium parameters depend arbitrarily on the coordinate $z$.
In this section, we restrict our attention mainly to the bilayer systems satisfying the matching conditions listed in Table~\ref{table1}.
Before presenting the numerical results obtained for those cases, we first consider a more generic case where the matching conditions are not satisfied to contrast the result with those in the matched cases.

\subsection{Generic case}

In Fig.~\ref{fig11} we plot the absorptances $A_1$ and $A_2$ for RCP and LCP waves of frequency $\omega$ incident on a bilayer system consisting of
a Tellegen medium with $\epsilon_1=2.25+0.01i$, $\mu_1=1$, $\chi_1=4$, and $\gamma_1=0$ and an ordinary dielectric
with $\epsilon_2=1.5$, $\mu_2=1$, and $\chi_2=\gamma_2=0$ in the configuration shown in Fig.~\ref{fig2} versus incident angle.
Plane waves are incident from a dielectric prism with the parameters $\epsilon_i=9$, $\mu_i=1$, and $a_i=0$.
The substrate is assumed to be made of the
same material as the prism. The layers 1 and 2 have the thicknesses $d_1=\Lambda$ and $d_2=4\Lambda$,
where $\Lambda$ satisfies $\omega\Lambda/c=0.2\pi$. For the given parameters,
we can solve Eq.~(\ref{eq:disp2}) numerically and find that the surface
wave is excited at the incident angle $\theta\approx 57.59^\circ$, which corresponds to $q\approx 2.5328$. This result agrees perfectly with the numerical result for the absorptances showing narrow sharp peaks at the same incident angle,
since the excitation of a surface wave is manifested by
a strong absorption of the energy of the incident wave. In the present example, it is interesting to notice that a surface wave
is excited even though the values of $\epsilon_1$, $\mu_1$, $\epsilon_2$, and $\mu_2$ are all positive, in contrast to the
more conventional cases where $\epsilon_1$ and $\epsilon_2$ (or $\mu_1$ and $\mu_2$) have the opposite signs.
This is possible because in the medium 1, the real part of the square of the effective refractive index ($=-13.75$) is negative
due to the large value
of the Tellegen parameter $\chi_1$. This makes the medium 1 behave similarly to a metal.

In Fig.~\ref{fig12} we show the spatial distributions of the intensities of the $y$ components
of the electric and magnetic fields, $\vert E_y\vert^2$ and $\vert H_y\vert^2$, when RCP and LCP waves
are incident on the same bilayer system considered in Fig.~\ref{fig11}.
The $E_y$ field associated with the incident wave is assumed to have a unit amplitude.
The incident angle is chosen to be $\theta=57^\circ$ corresponding to the angle at which the absorptances take the peak values.
We find that the field intensities are greatly enhanced at the interface between the two media and decay exponentially away from it, as can be expected from a surface wave.

\begin{figure}
\centering\includegraphics[width=8.5cm]{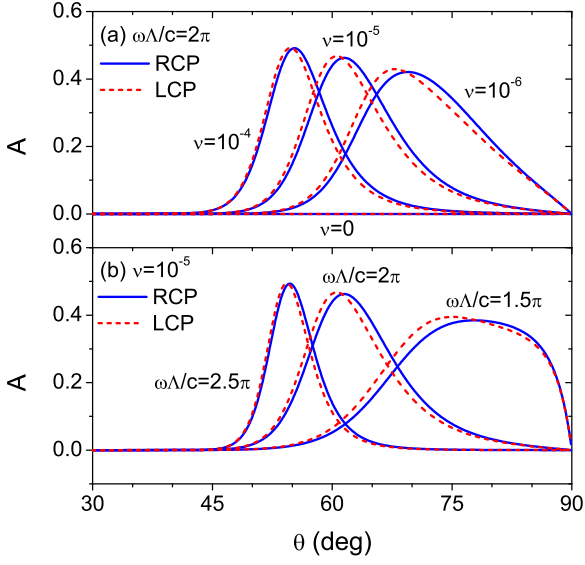}
\caption{Absorptances $A_1$ and $A_2$ for RCP and LCP waves of frequency $\omega$ incident on a bilayer system consisting of
two different kinds of Tellegen media with $\epsilon_1=-3+\nu i$, $\mu_1=-1$, and $\chi_1=-1$
and $\epsilon_2=3$, $\mu_2=1$, and $\chi_2=1$ in the configuration shown in Fig.~\ref{fig2} plotted versus incident angle.
Waves are incident from a prism with the parameters $\epsilon_i=4$, $\mu_i=1$, and $a_i=0$ and
transmitted to the substrate with the same parameters as the prism.
The layers 1 and 2 have the same thickness $\Lambda$ and the results for (a) four different values of $\nu$ ($=0$,
$10^{-4}$, $10^{-5}$, $10^{-6}$) when $\omega\Lambda/c=2\pi$ and three different values of $\omega\Lambda/c$ ($=1.5\pi$, $2\pi$,
$2.5\pi$) when $\nu=10^{-5}$ are shown.}
\label{fig21}
\end{figure}

\begin{figure}
\centering\includegraphics[width=8.5cm]{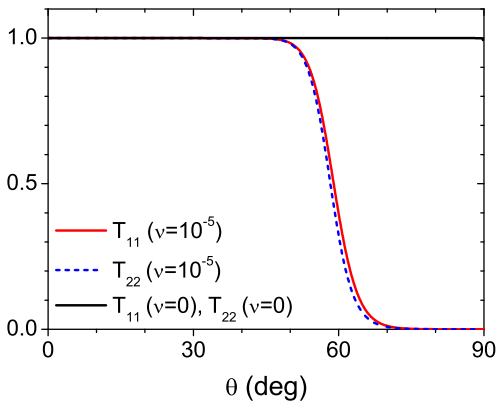}
\caption{Transmittances $T_{11}$ and $T_{22}$ for RCP and LCP waves of frequency $\omega$ in the same configuration as in Fig.~\ref{fig21}
plotted versus incident angle. The layers 1 and 2 have the same thickness $\Lambda$ such that $\omega\Lambda/c=2\pi$
and the results for two different values of $\nu$ ($=0$, $10^{-5}$) are compared.}
\label{nfig5}
\end{figure}

\begin{figure}
\centering\includegraphics[width=8.5cm]{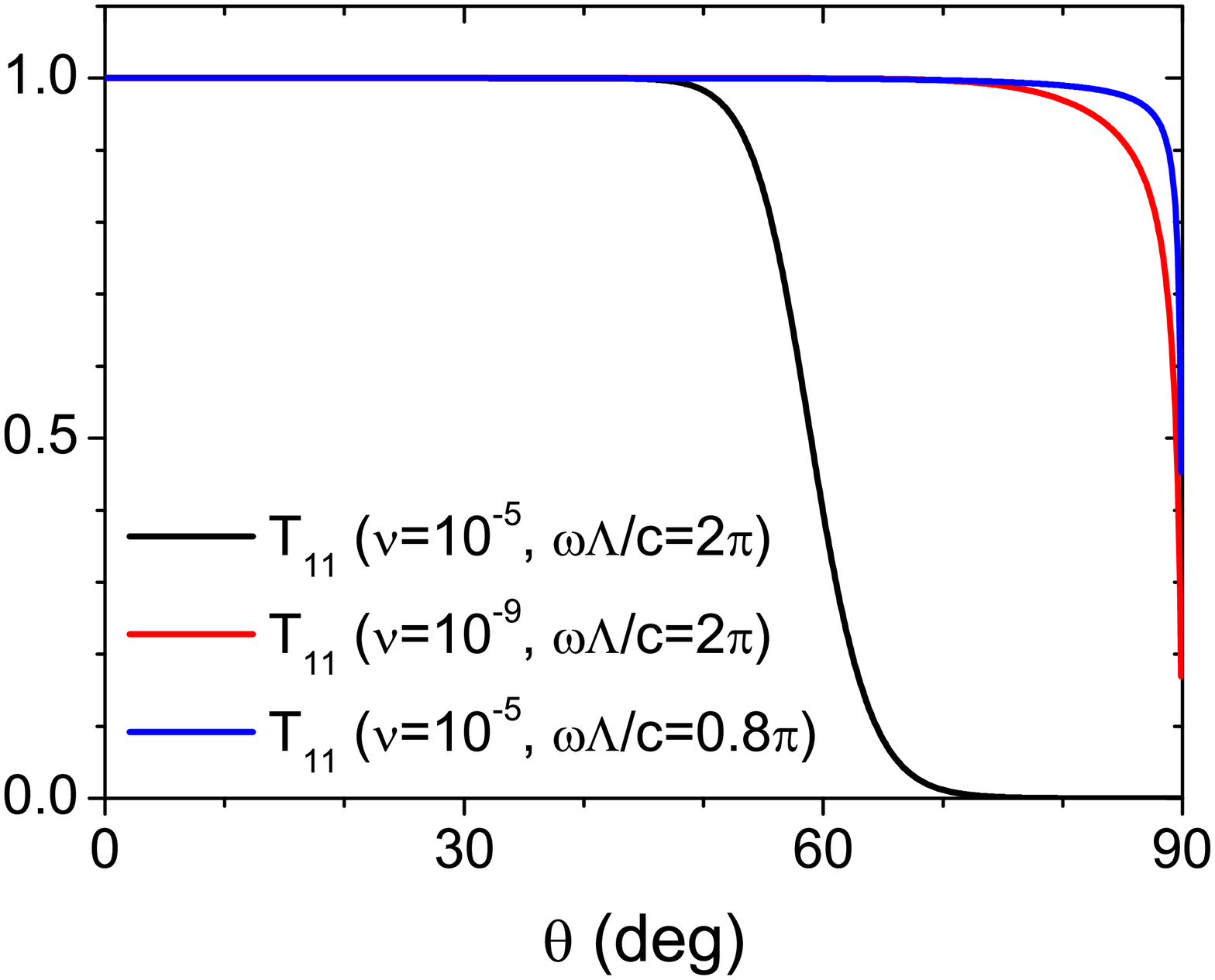}
\caption{Comparison of $T_{11}$ obtained for $\nu=10^{-5}$ and $\omega\Lambda/c=2\pi$ with those for
$\nu=10^{-9}$ and $\omega\Lambda/c=2\pi$ and for $\nu=10^{-5}$ and $\omega\Lambda/c=0.8\pi$
in the same configuration as in Fig.~\ref{fig21}.}
\label{newfig}
\end{figure}

\begin{figure}
\centering\includegraphics[width=8.5cm]{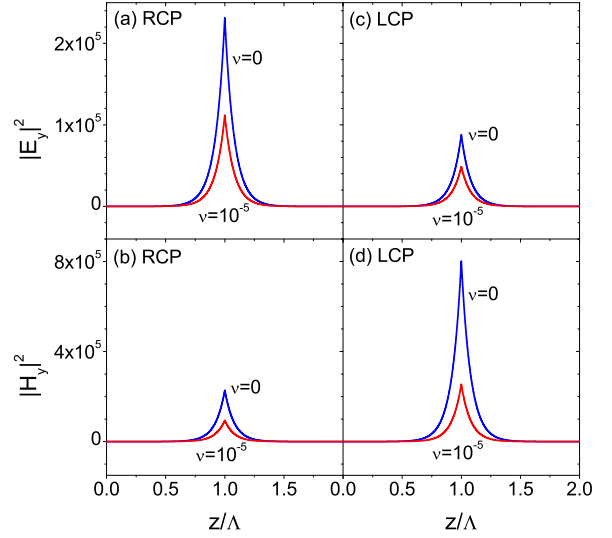}
\caption{Spatial distributions of the intensities of the $y$ components
of the electric and magnetic fields, $\vert E_y\vert^2$ and $\vert H_y\vert^2$, when RCP and LCP waves
with a unit amplitude of $E_y$
are incident on the same bilayer system considered in Fig.~\ref{nfig5}.
The incident angle is chosen to be $\theta=60^\circ$ and the parameter $\nu$ is either 0 or $10^{-5}$.
The interface is located at $z=\Lambda$ and waves are incident from the region where $z>2\Lambda$.}
\label{ffgg}
\end{figure}

\subsection{Cases I and II}

We now move to the cases where the matching conditions listed in Table~\ref{table1} are satisfied.
In Fig.~\ref{fig21} we consider the excitation of surface waves at the interface between two different Tellegen media (that is, media with $\chi\ne 0$ and $\gamma=0$) of the same thickness $\Lambda$,
where the medium parameters are given by $\epsilon_1=-3+\nu i$, $\mu_1=-1$, and
$\chi_1=-1$ and $\epsilon_2=3$, $\mu_2=1$, and
$\chi_2=1$ respectively. We also assume that waves are incident from a dielectric prism with the parameters
$\epsilon_i=4$, $\mu_i=1$, and $a_i=0$ and transmitted to the substrate with the same parameters as the prism.
If we ignore the small imaginary part in $\epsilon_1$ (that is, $\nu$), we obtain ${n_1}^2=\epsilon_1\mu_1-{\chi_1}^2=2$
and ${n_2}^2=\epsilon_2\mu_2-{\chi_2}^2=2$. When both $\epsilon$ and $\mu$ are negative and $\epsilon\mu>\chi^2$, we have to choose
$n=-\sqrt{\epsilon\mu-\chi^2}$ as the effective refractive index. Therefore we have $n_1=-n_2=-\sqrt{2}$ in the present case.
The effective impedances for RCP and LCP waves are obtained from Eq.~(\ref{eq:neta}) and we find that if we ignore $\nu$,
$\eta_{1r}=\eta_{2r}=(\sqrt{2}+i)/3$ and $\eta_{1l}=\eta_{2l}=(\sqrt{2}-i)/3$. When these conditions are satisfied, surface waves can be
excited for any real value of $\kappa_1$ ($=\kappa_2$), as we have proved in Eq.~(\ref{eq:cq1}).
Since a wave is incident from the region where $\sqrt{\epsilon_i\mu_i}=2$, this gives the
constraint that ${\kappa_1}^2/{k_0}^2=4\sin^2\theta-2>0$, that is, $\theta>45^\circ$ for $\kappa_1$ to be real. Therefore surface waves will be excited
for any value of $\theta$ greater than $45^\circ$ for both RCP and LCP incident waves. These predictions are confirmed clearly in Fig.~\ref{fig21}, where the absorptances $A_1$ and $A_2$ obtained for four different values of $\nu$ ($=0$,
$10^{-4}$, $10^{-5}$, and $10^{-6}$) when $\omega\Lambda/c=2\pi$ and three different values of $\omega\Lambda/c$ ($=1.5\pi$, $2\pi$, and
$2.5\pi$) when $\nu=10^{-5}$ are plotted versus incident angle. When $\nu$ is zero,
there is no dissipation in the layers and the absorption of the wave energy
does not arise, although even in this case, the surface waves are excited at the interface for $\theta>45^\circ$,
as will be shown in Fig.~\ref{ffgg}. When $\nu$ is small and nonzero, the absorptances remain zero at $\theta<45^\circ$, but
become finite at all angles greater than $45^\circ$. In the parameter region where the surface waves are excited, we find that
the absorptances depend quite
sensitively on the value of $\nu$ and $\omega\Lambda/c$. The broad peaks shown in Fig.~\ref{fig21} are in sharp contrast to the
narrow sharp peaks in Fig.~\ref{fig11}. In the calculations shown here and in all later calculations, we have assumed a very small
value of $\nu$. However, we have checked numerically that qualitatively similar results are
obtained for much larger values of $\nu$ up to $0.01$.

In the case where $\nu$ is zero, the bilayer system considered above is an example of a conjugate matched pair.
From Eq.~(\ref{eq:epmu}), we obtain $\epsilon_{2r}=-\epsilon_{1r}=2-\sqrt{2}i$, $\mu_{2r}=-\mu_{1r}=(2+\sqrt{2}i)/3$,
$\epsilon_{2l}=-\epsilon_{1l}=2+\sqrt{2}i$, and $\mu_{2l}=-\mu_{1l}=(2-\sqrt{2}i)/3$, which satisfy Eq.~(\ref{eq:cmp1}).
In Fig.~\ref{nfig5} we plot the transmittances $T_{11}$ and $T_{22}$ through the bilayer in the same configuration as in Fig.~\ref{fig21}.
As is expected from a conjugate matched pair, both $T_{11}$ and $T_{22}$ are identically equal to one for any incident angle when $\nu$ is zero.
For a small value of $\nu$ equal to $10^{-5}$, however, the transmittances remain equal to one only for $\theta<45^\circ$.
This occurs because when the surface waves are excited at $\theta>45^\circ$, a large amount of absorption arises even in the presence of
a very small damping.
In all cases where the omnidirectional total transmission occurs, the cross-polarized transmittances $T_{12}$ and $T_{21}$
obviously vanish. In other cases, however, they are generally nonzero.

In Fig.~\ref{newfig}, we make a comparison of the transmittance $T_{11}$ obtained for $\nu=10^{-5}$ and $\omega\Lambda/c=2\pi$ with those for
$\nu=10^{-9}$ and $\omega\Lambda/c=2\pi$ and for $\nu=10^{-5}$ and $\omega\Lambda/c=0.8\pi$
in the same configuration as in Fig.~\ref{fig21}. Although true omnidirectional total transmission occurs only when $\nu=0$, we find that
total transmission is obtained at all angles except for those very close to $90^\circ$, if $\nu$ or $\Lambda$ is sufficiently small.
Tuning the slab thickness $\Lambda$ seems to be a convenient way to observe both the omnidirectional total transmission
and the omnidirectional excitation of surface waves experimentally.

In Fig.~\ref{ffgg}, we show the spatial distributions of $\vert E_y\vert^2$ and $\vert H_y\vert^2$, when RCP and LCP waves
with a unit amplitude of $E_y$ are incident on the same bilayer system considered in Fig.~\ref{nfig5}
at the incident angle $\theta=60^\circ$ for two different values of $\nu$ ($=0$, $10^{-5}$).
We find that both fields, which are extremely pronounced near the interface between the two media, decay exponentially away from it.
We have checked numerically that
the decay rates of the fields are approximately consistent with the values obtained from $\kappa_1$ and $\kappa_2$ for the chosen value of $\theta$.
We observe that the surface waves are excited even when $\nu$ is zero and the field enhancement in that case is larger than that in the $\nu\ne 0$ case. We emphasize that these are the solutions in the steady state.

\begin{figure}
\centering\includegraphics[width=8.5cm]{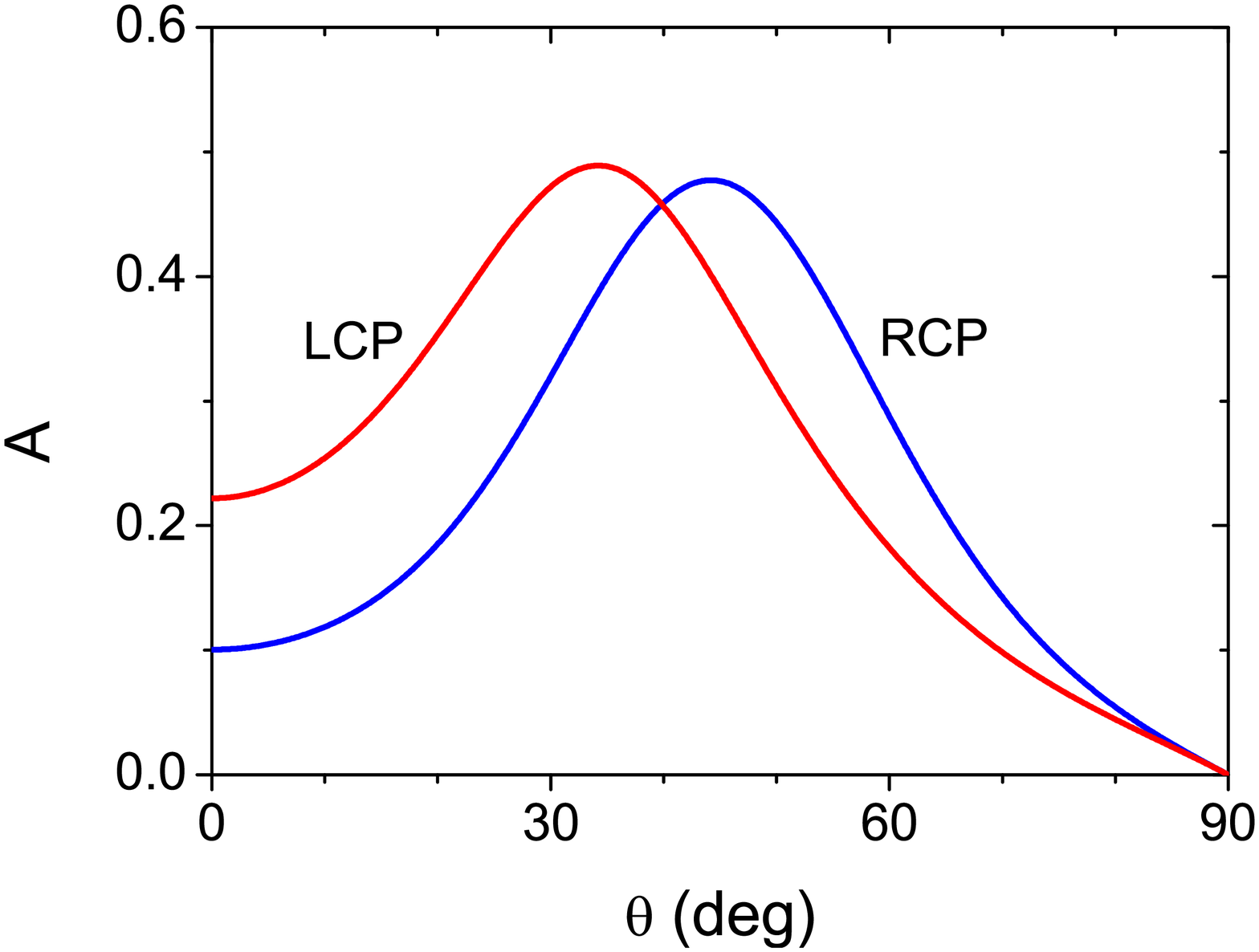}
\caption{Absorptances $A_1$ and $A_2$ for RCP and LCP waves of frequency $\omega$ incident on a bilayer system consisting of
two different kinds of Tellegen media with $\epsilon_1=-3+\nu i$ ($\nu=10^{-5}$), $\mu_1=1$, and $\chi_1=1$
and $\epsilon_2=3$, $\mu_2=-1$, and $\chi_2=-1$ plotted versus incident angle.
Waves are incident from a prism with the parameters $\epsilon_i=4$, $\mu_i=1$, and $a_i=0$ and
transmitted to the substrate with the same parameters as the prism.
The layers 1 and 2 have the same thickness $\Lambda$, which satisfies $\omega\Lambda/c=\pi$.}
\label{fig31}
\end{figure}

\begin{figure}
\centering\includegraphics[width=8.5cm]{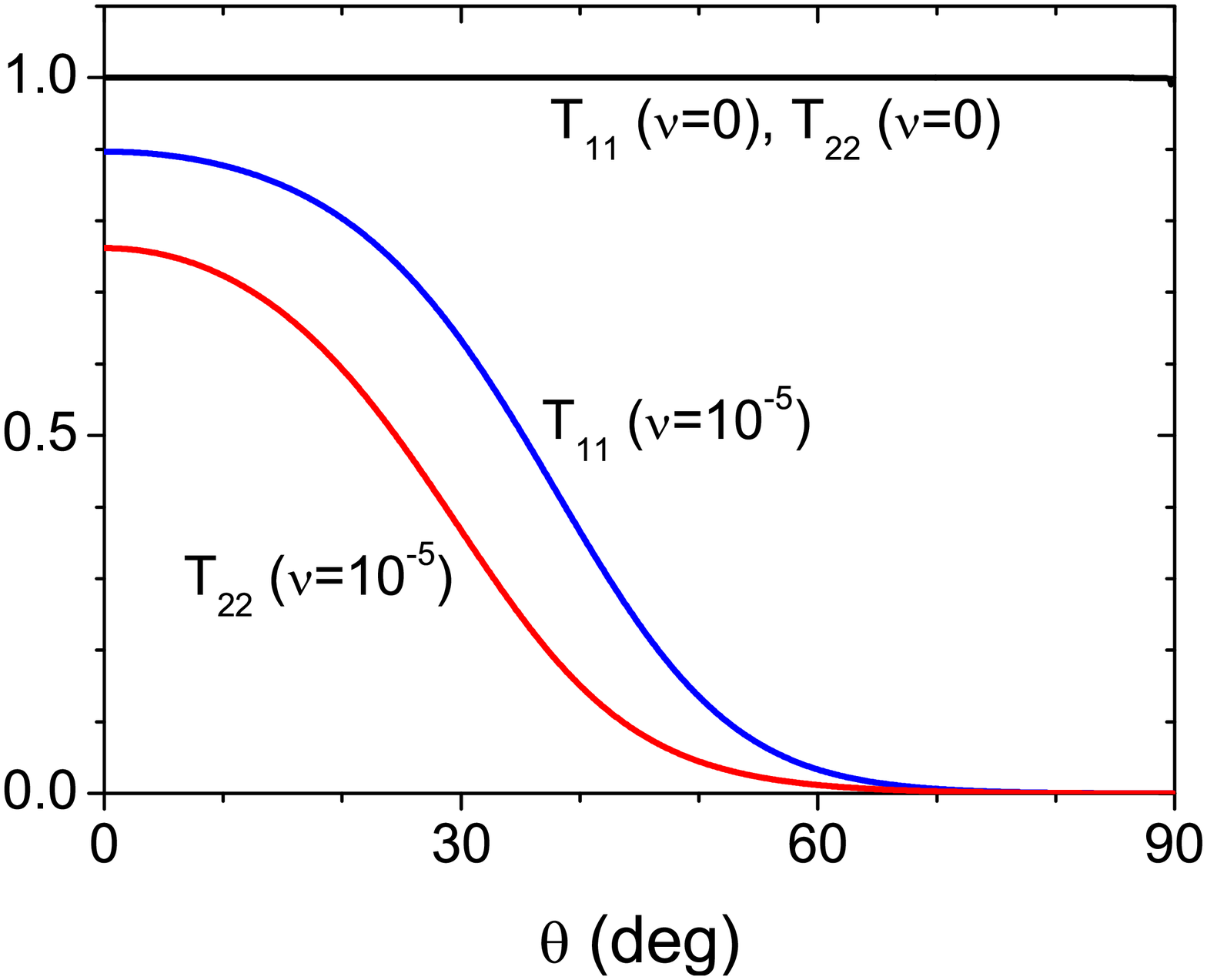}
\caption{Transmittances $T_{11}$ and $T_{22}$ for RCP and LCP waves in the same configuration as in Fig.~\ref{fig31}
plotted versus incident angle when $\nu$ is 0 and $10^{-5}$.}
\label{nfig7}
\end{figure}

\begin{figure}
\centering\includegraphics[width=8.5cm]{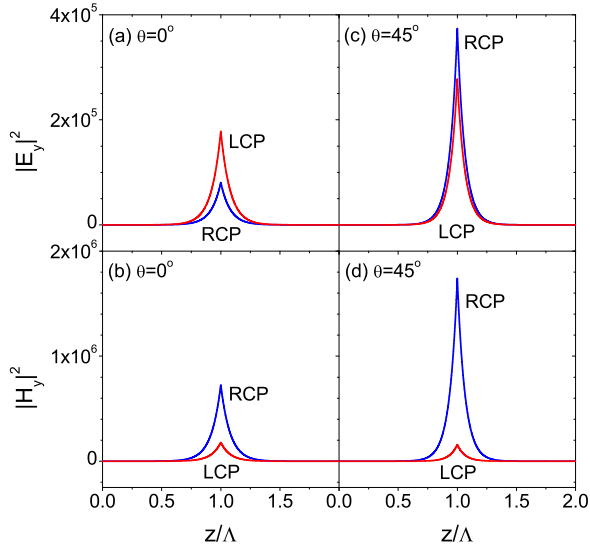}
\caption{Spatial distributions of the intensities of the $y$ components
of the electric and magnetic fields, $\vert E_y\vert^2$ and $\vert H_y\vert^2$, when RCP and LCP waves
with a unit amplitude of $E_y$
are incident on the same bilayer system considered in Fig.~\ref{fig31}.
The incident angle is chosen to be either $0^\circ$ or $45^\circ$.
The interface is located at $z=\Lambda$ and waves are incident from the region where $z>2\Lambda$.}
\label{fig32}
\end{figure}

\subsection{Cases III and IV}

Next, in Fig.~\ref{fig31} we consider the surface wave between two Tellegen media of the same thickness $\Lambda$ such that $\omega\Lambda/c=\pi$,
where $\epsilon_1=-3+\nu i$ ($\nu=10^{-5}$), $\mu_1=1$, and
$\chi_1=1$ and $\epsilon_2=3$, $\mu_2=-1$, and
$\chi_2=-1$ respectively. We also assume that waves are incident from a dielectric prism with the parameters
$\epsilon_i=4$, $\mu_i=1$, and $a_i=0$ and transmitted to the substrate with the same parameters as the prism.
If we ignore $\nu$, we obtain $n_1=n_2=2i$, $\eta_{1r}=-\eta_{2l}=-i$, and
$\eta_{1l}=-\eta_{2r}=-i/3$.
When these conditions are satisfied, surface waves can be
excited for any real value of $\kappa_1$ ($=\kappa_2$) for both RCP and LCP incident waves,
as we have proved in Eq.~(\ref{eq:cq2}). Since a wave is incident from the region where $\sqrt{\epsilon_i\mu_i}=2$, this gives the
condition that ${\kappa_1}^2/{k_0}^2=4\sin^2\theta+2>0$, which is satisfied for all $\theta$. Therefore surface waves will be excited
for an arbitrary value of $\theta$ including the normal incidence case with $\theta=0$, which is clearly confirmed in Fig.~\ref{fig31}.

We can also show that when $\nu$ is zero, the bilayer system considered in Fig.~\ref{fig31} is a conjugate matched pair.
From Eq.~(\ref{eq:epmu}), we obtain $\epsilon_{2r}=-\epsilon_{1l}=6$, $\mu_{2r}=-\mu_{1l}=-2/3$,
$\epsilon_{2l}=-\epsilon_{1r}=2$, and $\mu_{2l}=-\mu_{1r}=-2$, which satisfy Eq.~(\ref{eq:cmp2}).
In Fig.~\ref{nfig7} we plot the transmittances $T_{11}$ and $T_{22}$ through the bilayer in the same configuration as in Fig.~\ref{fig31}.
When $\nu$ is zero, we confirm that the omnidirectional total transmission of both RCP and LCP waves occurs.
When $\nu$ is nonzero, the transmittances are substantially reduced at all $\theta$ due to the omnidirectional excitation of surface waves.

In Fig.~\ref{fig32}, we show the spatial distributions of $\vert E_y\vert^2$ and $\vert H_y\vert^2$, when RCP and LCP waves
with a unit amplitude of $E_y$ are incident on the same bilayer system considered in Fig.~\ref{fig31} with $\nu=10^{-5}$
at the angles $\theta=0^\circ$ and $45^\circ$.
In all cases including the normal incidence case,
we find that surface waves are strongly excited at the interface.

\begin{figure}
\centering\includegraphics[width=8.5cm]{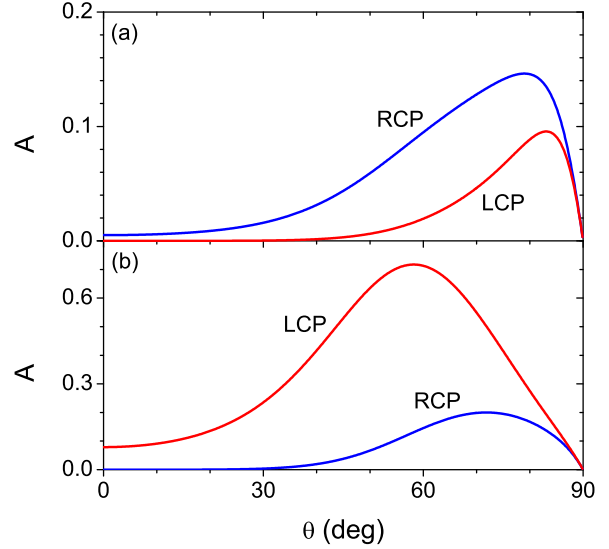}
\caption{Absorptances $A_1$ and $A_2$ for RCP and LCP waves of frequency $\omega$ incident on a bilayer system consisting of
two different kinds of Tellegen media (a) with $\epsilon_1=-3+\nu i$ ($\nu=10^{-5}$), $\mu_1=-7$, and $\chi_1=5$
and $\epsilon_2=-3$, $\mu_2=1$, and $\chi_2=1$ and (b) with $\epsilon_1=-3+\nu i$ ($\nu=10^{-5}$), $\mu_1=1$, and $\chi_1=1$
and $\epsilon_2=-3$, $\mu_2=-7$, and $\chi_2=5$ plotted versus incident angle.
Waves are incident from a prism with the parameters $\epsilon_i=4$, $\mu_i=1$, and $a_i=0$ and
transmitted to the substrate with the same parameters as the prism.
The layers 1 and 2 have the same thickness $\Lambda$, which satisfies $\omega\Lambda/c=0.8\pi$.}
\label{fig41}
\end{figure}

\begin{figure}
\centering\includegraphics[width=8.5cm]{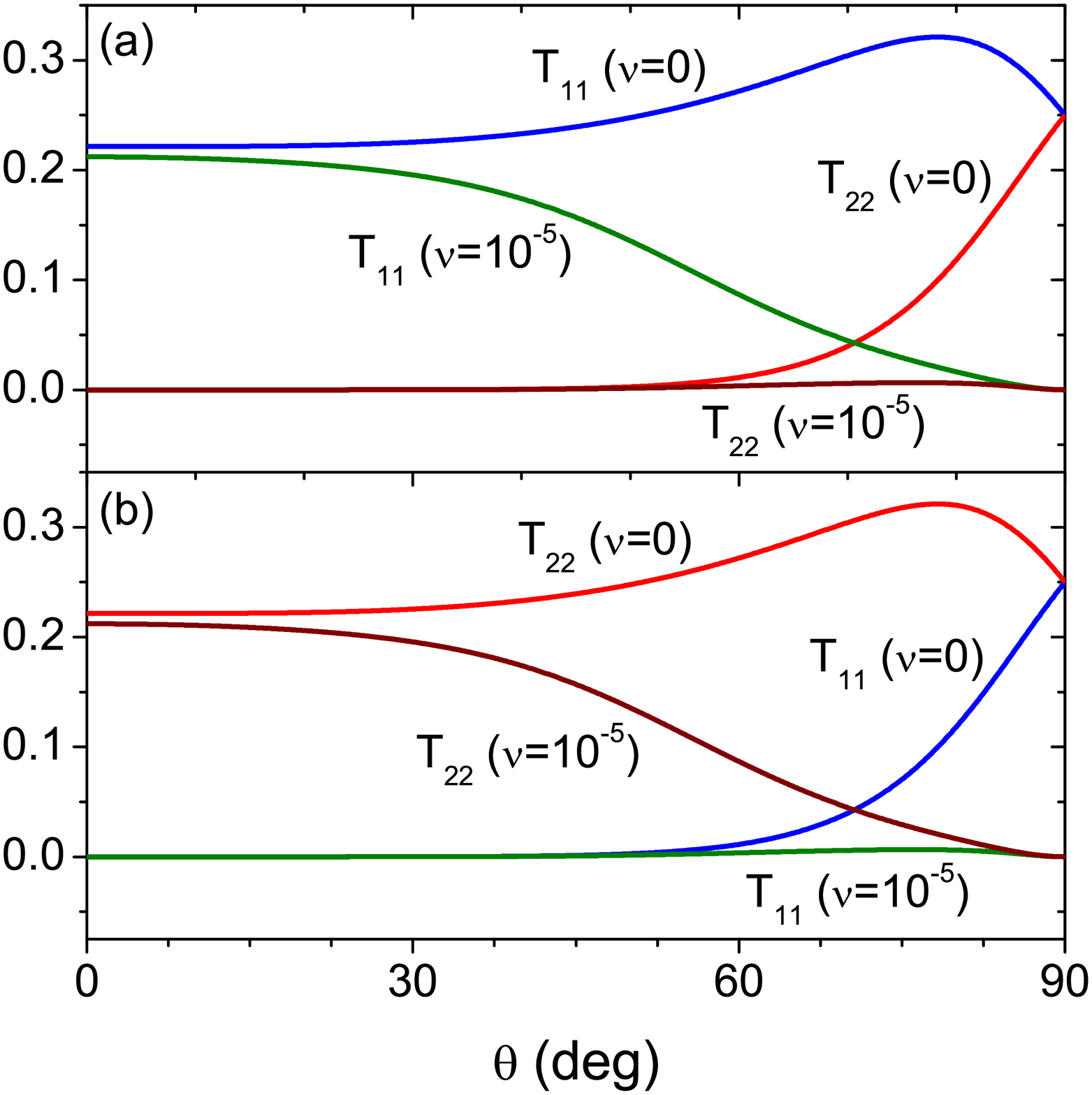}
\caption{Transmittances $T_{11}$ and $T_{22}$ for RCP and LCP waves in the same configurations as in Fig.~\ref{fig41}
plotted versus incident angle when $\nu$ is 0 and $10^{-5}$.}
\label{nfig10a}
\end{figure}

\subsection{Case III or IV}
\label{ss_a}

We now consider the third configuration involving two Tellegen media.
In Fig.~\ref{fig41}(a) we consider the surface wave between two Tellegen media of the same thickness $\Lambda$
such that $\omega\Lambda/c=0.8\pi$,
where $\epsilon_1=-3+\nu i$ ($\nu=10^{-5}$), $\mu_1=-7$, and
$\chi_1=5$ and $\epsilon_2=-3$, $\mu_2=1$, and
$\chi_2=1$ respectively. We assume that waves are incident from a dielectric prism with the parameters
$\epsilon_i=4$, $\mu_i=1$, and $a_i=0$ and transmitted to the substrate with the same parameters as the prism.
If we ignore $\nu$, we obtain $n_1=n_2=2i$ and
the effective impedances for RCP and LCP waves
$\eta_{1r}=-7i/3$, $\eta_{1l}=i$, $\eta_{2r}=-i$, and $\eta_{2l}=-i/3$, and therefore we have $\eta_{1l}=-\eta_{2r}$
but $\eta_{1r}\ne -\eta_{2l}$. In this case, we find that surface waves are
excited for an arbitrary value of $\theta$ for RCP incident waves. For LCP waves, surfaces waves are also excited at
any incident angle greater than 0, but they are not excited at $\theta=0$ and very weakly excited at small incident angles.

These results can be understood from the polarization of the surface-wave mode, $(0,1,1,0)$, corresponding to the
fourth case in Table~\ref{table1}. This implies that the surface wave in this case is LCP in the medium 1 and RCP in the medium 2.
Let us first consider the normal incidence case. When an RCP (LCP) wave is normally incident on the interface between the two media, the transmitted wave
will be RCP (LCP) and the reflected wave will be LCP (RCP). Therefore it is not possible to have the correct mode structure
when an LCP wave is incident and the surface wave will not be excited in that case. For a nonzero incident angle, the reflected
and transmitted waves will have both RCP and LCP components. When $\theta$ is sufficiently small, however,
the coupling and mixing between RCP and LCP components is weak and the behavior is similar to the normal incidence case.
In Fig.~\ref{fig41}(b) we have switched the parameters of the media 1 and 2 (except for $\nu$). Then we obtain $n_1=n_2$, $\eta_{1r}=-\eta_{2l}$, and
$\eta_{1l}\ne -\eta_{2r}$. A similar argument as the above shows that surface waves are
excited at all $\theta$ for LCP incident waves. For RCP waves, however, surface waves are not excited at $\theta=0$ and very weakly excited at small incident angles, as is confirmed in Fig.~\ref{fig41}(b).

If we ignore $\nu$, we also obtain $\epsilon_{2r}=-\epsilon_{1l}=-2$, $\mu_{2r}=-\mu_{1l}=2$,
$\epsilon_{2l}=-6$, $\epsilon_{1r}=-6/7$, $\mu_{2l}=2/3$, $\mu_{1r}=14/3$ from Eq.~(\ref{eq:epmu}).
Therefore the second condition of Eq.~(\ref{eq:cmp2}) is satisfied but the first is not, and therefore this bilayer is not a conjugate
matched pair.
In Fig.~\ref{nfig10a} we plot the transmittances $T_{11}$ and $T_{22}$ through the bilayer in the same configuration as in Fig.~\ref{fig41}.
We confirm that the omnidirectional total transmission does not occur regardless of the value of $\nu$.

\begin{figure}
\centering\includegraphics[width=8.5cm]{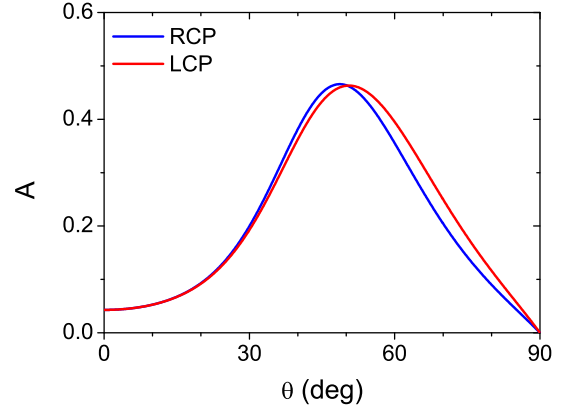}
\caption{Absorptances $A_1$ and $A_2$ for RCP and LCP waves of frequency $\omega$ incident on a bilayer system consisting of
two different kinds of chiral media with $\epsilon_1=-3+\nu i$ ($\nu=10^{-5}$), $\mu_1=1$, and $\gamma_1=0.1$
and $\epsilon_2=3$, $\mu_2=-1$, and $\gamma_2=-0.1$ in the configuration shown in Fig.~\ref{fig2} plotted versus incident angle.
Waves are incident from a prism with the parameters $\epsilon_i=4$, $\mu_i=1$, and $a_i=0$ and
transmitted to the substrate with the same parameters as the prism.
The layers 1 and 2 have the same thickness $\Lambda$, which satisfies $\omega\Lambda/c=\pi$.}
\label{fig51}
\end{figure}

\begin{figure}
\centering\includegraphics[width=8.5cm]{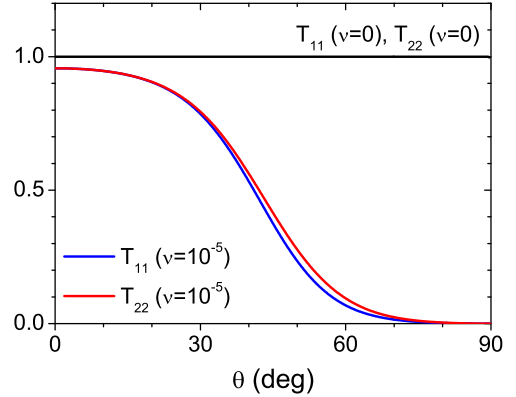}
\caption{Transmittances $T_{11}$ and $T_{22}$ for RCP and LCP waves in the same configuration as in Fig.~\ref{fig51}
plotted versus incident angle when $\nu$ is 0 and $10^{-5}$.}
\label{nfig11}
\end{figure}

\subsection{Cases V and VI}

Next, we consider the excitation of surface waves at the interface between two different chiral media
(that is, media with $\gamma\ne 0$ and $\chi=0$),
where $\eta_{jr}=\eta_{jl}=\eta_j$ ($j=1,2$).
In Fig.~\ref{fig51} we consider the case where $\epsilon_1=-3+\nu i$ ($\nu=10^{-5}$), $\mu_1=1$, and $\gamma_1=0.1$
and $\epsilon_2=3$, $\mu_2=-1$, and $\gamma_2=-0.1$. The two layers have the same thickness $\Lambda$
such that $\omega\Lambda/c=\pi$. If we ignore $\nu$,
we obtain $\eta_1=-\eta_2=-i/\sqrt{3}$,
$n_{1r}=n_{2l}=0.1+\sqrt{3}i$, and $n_{1l}=n_{2r}=-0.1+\sqrt{3}i$.
In this case, we find that surface waves are
excited for an arbitrary value of $\theta$ for both RCP and LCP incident waves, as we have proved in Eq.~(\ref{eq:cq3})
and verified in Fig.~\ref{fig51}.

We find that when $\nu$ is zero, the bilayer considered in Fig.~\ref{fig51} is a conjugate matched pair.
From Eq.~(\ref{eq:epmu}), we obtain $\epsilon_{2r}=-\epsilon_{1l}=3+0.1\sqrt{3}i$, $\mu_{2r}=-\mu_{1l}=-1-(0.1/\sqrt{3})i$,
$\epsilon_{2l}=-\epsilon_{1r}=3-0.1\sqrt{3}i$, and $\mu_{2l}=-\mu_{1r}=-1+(0.1/\sqrt{3})i$, which satisfy Eq.~(\ref{eq:cmp2}).
In Fig.~\ref{nfig11} we plot the transmittances $T_{11}$ and $T_{22}$ through the bilayer in the same configuration as in Fig.~\ref{fig51}
and confirm that the omnidirectional total transmission of both RCP and LCP waves occurs when $\nu$ is zero.
When $\nu$ is nonzero, the transmittances are reduced at all $\theta$ due to the omnidirectional excitation of surface waves.

\begin{figure}
\centering\includegraphics[width=8.5cm]{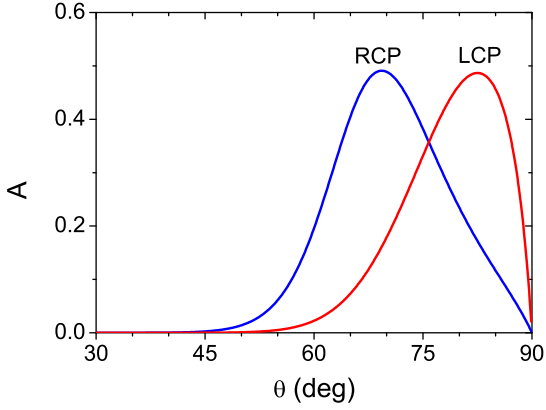}
\caption{Absorptances $A_1$ and $A_2$ for RCP and LCP waves of frequency $\omega$ incident on a bilayer system consisting of
two different kinds of chiral media with $\epsilon_1=-1.4+\nu i$ ($\nu=10^{-5}$), $\mu_1=-1.4$, and $\gamma_1=0.1$
and $\epsilon_2=1.4$, $\mu_2=1.4$, and $\gamma_2=-0.1$ plotted versus incident angle.
In the incident region and the substrate, the medium parameters are $\epsilon_i=\epsilon_t=2$, $\mu_i=\mu_t=2$, and $a_i=a_t=0$.
The layers 1 and 2 have the same thickness $\Lambda$, which satisfies $\omega\Lambda/c=1.5\pi$.}
\label{fig61}
\end{figure}

\begin{figure}
\centering\includegraphics[width=8.5cm]{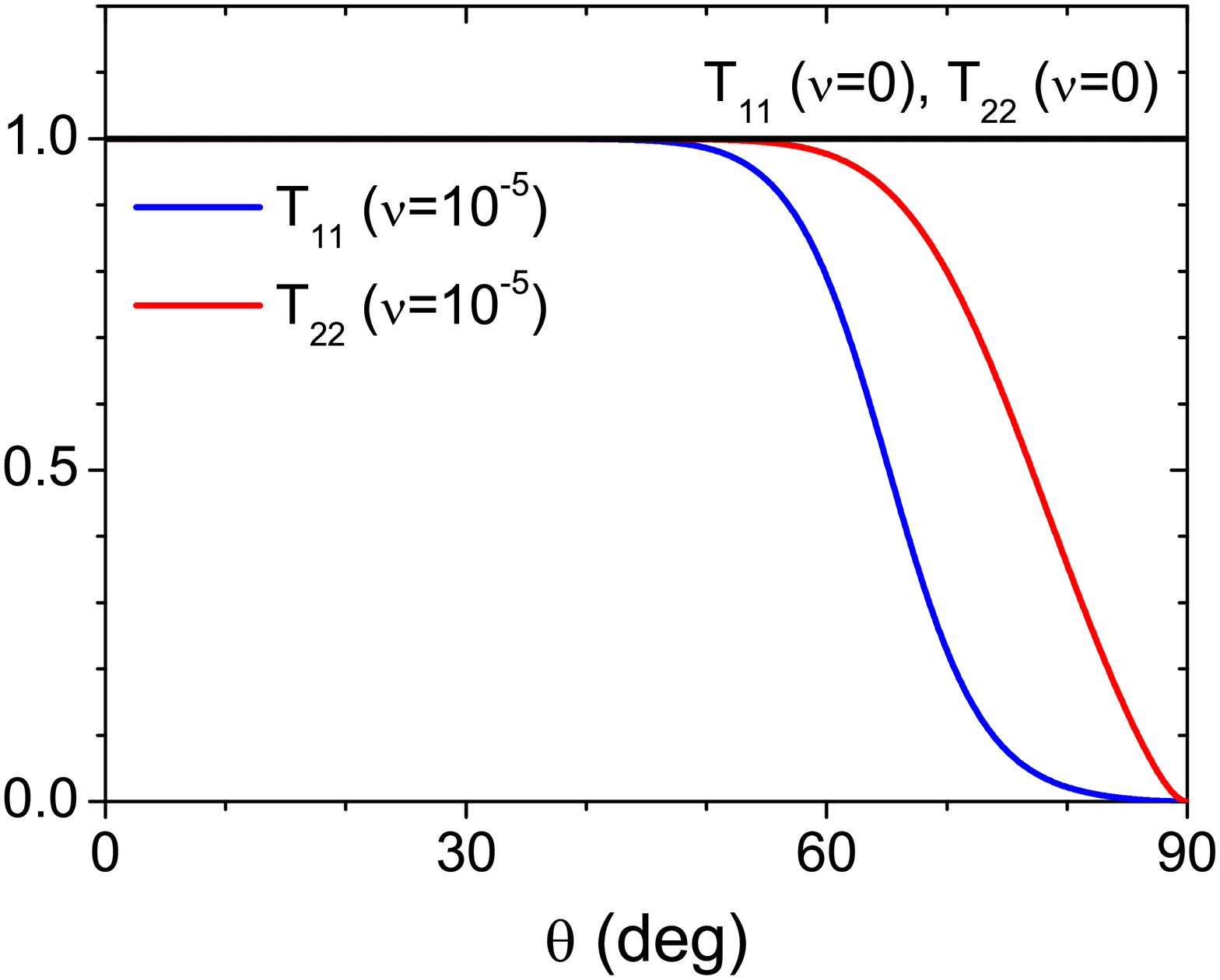}
\caption{Transmittances $T_{11}$ and $T_{22}$ for RCP and LCP waves in the same configuration as in Fig.~\ref{fig61}
plotted versus incident angle when $\nu$ is 0 and $10^{-5}$.}
\label{nfig13}
\end{figure}

\subsection{Cases VII and VIII}

From now on, we will consider the cases where the effective impedances of the media 1 and 2 are the same
when we ignore the small imaginary part in $\epsilon_1$.
In those cases, it is more illuminating to have the same impedance value in both the incident region and the substrate as in the media 1 and 2.
Then the impedance is matched throughout the whole space and the unwanted wave scattering
at the interfaces between the incident region and the medium 1 and between the medium 2 and the substrate is eliminated.

In Fig.~\ref{fig61} we consider the case where $\epsilon_1=-1.4+\nu i$ ($\nu=10^{-5}$), $\mu_1=-1.4$, and $\gamma_1=0.1$
and $\epsilon_2=1.2$, $\mu_2=1.2$, and $\gamma_2=0.1$. The layers 1 and 2 have the same thickness $\Lambda$, which satisfies $\omega\Lambda/c=1.5\pi$.
In the incident region and the substrate, we choose $\epsilon_i=\epsilon_t=2$, $\mu_i=\mu_t=2$, and $a_i=a_t=0$
to have $\eta_i=\eta_t=1$. If we ignore $\nu$,
we obtain $\eta_1=\eta_2=\eta_i=\eta_t=1$.
Ignoring $\nu$, the effective refractive indices for RCP and LCP waves are given by
$n_{1r}=-1.3$, $n_{1l}=-1.5$, $n_{2r}=1.3$, and $n_{2l}=1.5$, and therefore we have $n_{1r}=-n_{2r}$ and
$n_{1l}=-n_{2l}$. When these conditions are satisfied, surface waves can be
excited for any real value of $\kappa_{1r}$ ($\kappa_{1l}$) for incident RCP (LCP) waves, as
we have proved in Eq.~(\ref{eq:cq4}). Since a wave is incident from the region where $\sqrt{\epsilon_i\mu_i}=2$, this gives the
constraints that ${\kappa_{1r}}^2/{k_0}^2=4\sin^2\theta-1.3^2>0$, that is, $\theta>40.54^\circ$ for RCP waves and ${\kappa_{1l}}^2/{k_0}^2=4\sin^2\theta-1.5^2>0$, that is, $\theta>48.59^\circ$ for LCP waves.
That the surface waves are indeed excited in the predicted regions of the incident angle
can be seen clearly from Fig.~\ref{fig61}.

When $\nu$ is zero, the bilayer considered in Fig.~\ref{fig61} is a conjugate matched pair.
From Eq.~(\ref{eq:epmu}), we obtain $\epsilon_{2r}=-\epsilon_{1r}=1.3$, $\mu_{2r}=-\mu_{1r}=1.3$,
$\epsilon_{2l}=-\epsilon_{1l}=1.5$, and $\mu_{2l}=-\mu_{1l}=1.5$, which satisfy Eq.~(\ref{eq:cmp1}).
In Fig.~\ref{nfig13} we plot the transmittances $T_{11}$ and $T_{22}$ through the bilayer in the same configuration as in Fig.~\ref{fig61}
and confirm that the omnidirectional total transmission of both RCP and LCP waves occurs when $\nu$ is zero.
For a small value of $\nu$ equal to $10^{-5}$, the transmittance remains equal to one only at $\theta<40.54^\circ$ for RCP waves
and at $\theta<48.59^\circ$ for LCP waves.
This is because the surface waves are excited at either $\theta>40.54^\circ$ or $\theta>48.59^\circ$ depending on the polarization of the incident wave.

In the present example, having the impedance matching in the whole space is not required to have the omnidirectional total transmission,
since the two conditions in Eq.~(\ref{eq:cmp1}) are simultaneously satisfied.
We have checked numerically that even in the case where $\epsilon_i=\epsilon_t=4$, $\mu_i=\mu_t=1$, and $a_i=a_t=0$,
the transmittances $T_{11}$ and $T_{22}$ are identically equal to one at all $\theta$ if $\nu$ is zero.
When $\nu$ is small and nonzero, the surface waves are excited at $\theta>40.54^\circ$ for both RCP and LCP waves and
the absorptances are slightly different from those shown in Fig.~\ref{fig61}
due to the coupling and mixing of RCP and LCP waves.

\begin{figure}
\centering\includegraphics[width=8.5cm]{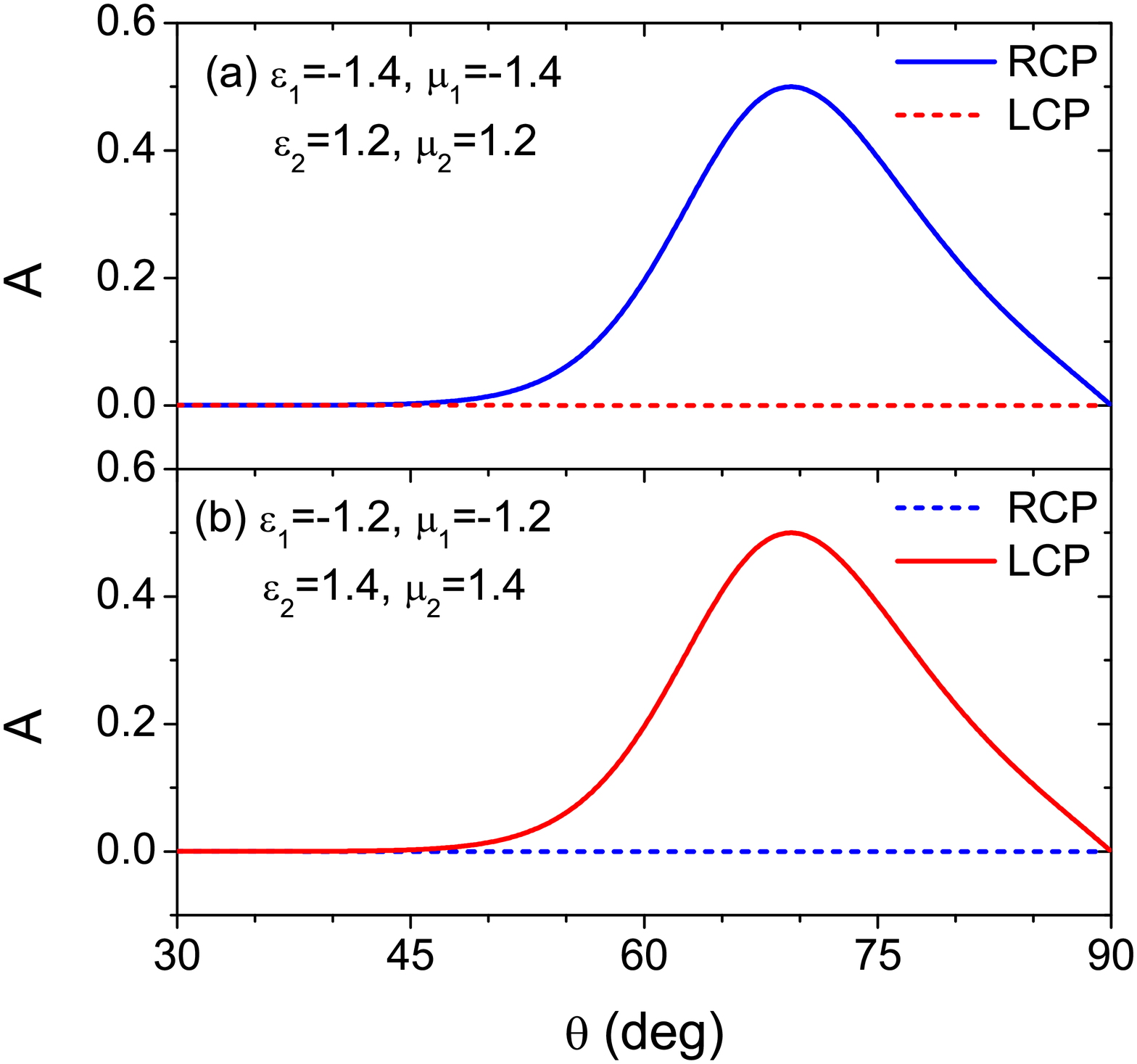}
\caption{Absorptances $A_1$ and $A_2$ for RCP and LCP waves of frequency $\omega$ incident on a bilayer system consisting of
two different kinds of chiral media (a) with $\epsilon_1=-1.4+\nu i$ ($\nu=10^{-5}$), $\mu_1=-1.4$, and $\gamma_1=0.1$
and $\epsilon_2=1.2$, $\mu_2=1.2$, and $\gamma_2=0.1$ and (b) with $\epsilon_1=-1.2+\nu i$ ($\nu=10^{-5}$), $\mu_1=-1.2$, and $\gamma_1=0.1$
and $\epsilon_2=1.4$, $\mu_2=1.4$, and $\gamma_2=0.1$ plotted versus incident angle.
In the incident region and the substrate, the medium parameters are $\epsilon_i=\epsilon_t=2$, $\mu_i=\mu_t=2$, and $a_i=a_t=0$.
The layers 1 and 2 have the same thickness $\Lambda$, which satisfies $\omega\Lambda/c=1.5\pi$.}
\label{fig71b}
\end{figure}

\begin{figure}
\centering\includegraphics[width=8.5cm]{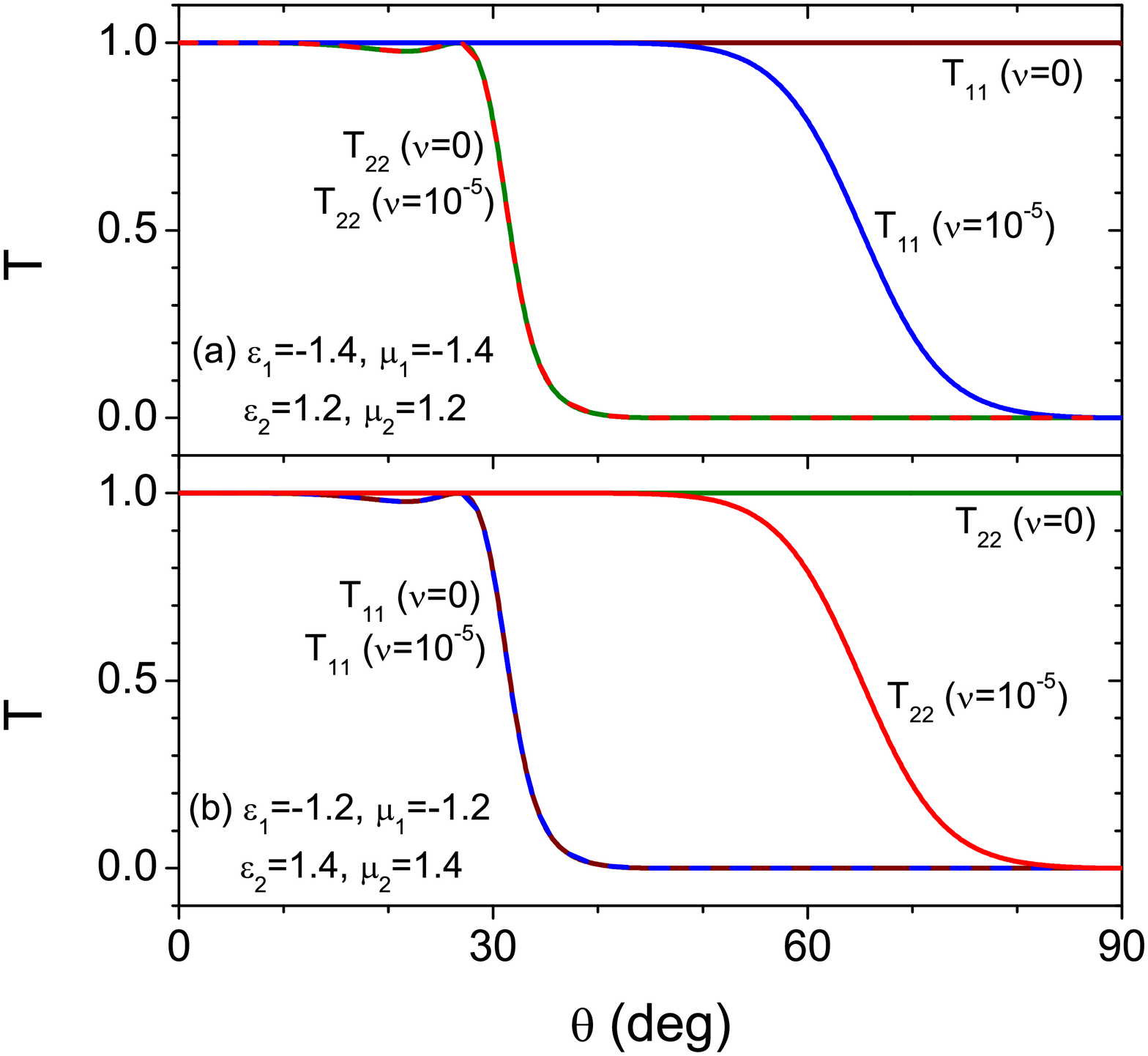}
\caption{Transmittances $T_{11}$ and $T_{22}$ for RCP and LCP waves in the same configurations as in Fig.~\ref{fig71b}
plotted versus incident angle when $\nu$ is 0 and $10^{-5}$.}
\label{fig71}
\end{figure}

\begin{figure}
\centering\includegraphics[width=8.5cm]{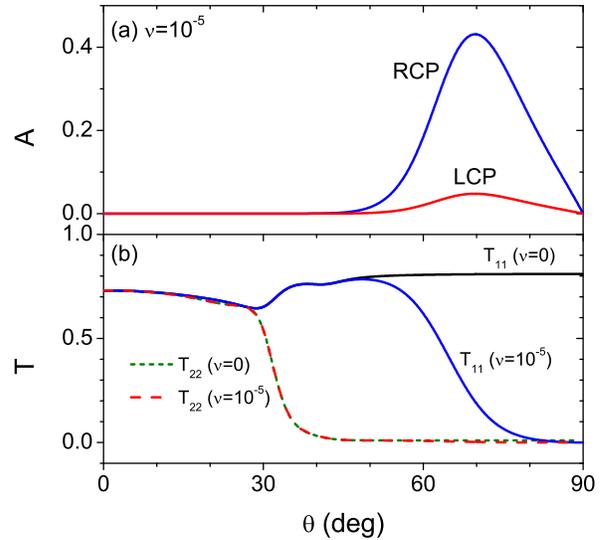}
\caption{(a) Absorptances $A_1$ and $A_2$ for RCP and LCP waves incident on the same bilayer considered in Fig.~\ref{fig71b}(a) plotted versus incident angle, when
the medium parameters in the incident region and the substrate are $\epsilon_i=\epsilon_t=4$, $\mu_i=\mu_t=1$, and $a_i=a_t=0$.
(b) Transmittances $T_{11}$ and $T_{22}$ for RCP and LCP waves in the same configuration as in Fig.~\ref{fig71}(a)
plotted versus incident angle when $\nu$ is 0 and $10^{-5}$.}
\label{fig19}
\end{figure}

\subsection{Case VII or VIII}

Finally, we consider the cases which satisfy Eq.~(\ref{eq:cmp3}).
In Fig.~\ref{fig71b}(a) we consider the case where $\epsilon_1=-1.4+\nu i$ ($\nu=10^{-5}$), $\mu_1=-1.4$, and $\gamma_1=0.1$
and $\epsilon_2=1.2$, $\mu_2=1.2$, and $\gamma_2=0.1$. In the incident region and the substrate, the medium parameters are $\epsilon_i=\epsilon_t=2$, $\mu_i=\mu_t=2$, and $a_i=a_t=0$.
The layers 1 and 2 have the same thickness $\Lambda$, which satisfies $\omega\Lambda/c=1.5\pi$.
If we ignore $\nu$, we obtain $\eta_1=\eta_2=\eta_i=\eta_t=1$.
Ignoring $\nu$, the effective refractive indices for RCP and LCP waves are given by
$n_{1r}=-1.3$, $n_{1l}=-1.5$, $n_{2r}=1.3$, and $n_{2l}=1.1$, and therefore we have $n_{1r}=-n_{2r}$ but
$n_{1l}\ne -n_{2l}$. When these conditions are satisfied, surface waves can be
excited for any real value of $\kappa_{1r}$ only for RCP waves as stated in Eq.~(\ref{eq:cmp3}). Since a wave is incident from the region where $\sqrt{\epsilon_i\mu_i}=2$, this gives the
constraint that $\theta>40.54^\circ$ for RCP waves.
We confirm that the surface waves are excited in the predicted region of the incident angle only when RCP waves are incident
in Fig.~\ref{fig71b}(a).
In Fig.~\ref{fig71b}(b) we consider the case where $\epsilon_1=-1.2+\nu i$ ($\nu=10^{-5}$), $\mu_1=-1.2$, and $\gamma_1=0.1$
and $\epsilon_2=1.4$, $\mu_2=1.4$, and $\gamma_2=0.1$. In this case, it is straightforward to verify that the surface waves are excited
when $\theta>40.54^\circ$ only for LCP incident waves.

When $\nu$ is zero, the bilayer system considered in Fig.~\ref{fig71b}(a) is a conjugate matched pair for RCP waves.
From Eq.~(\ref{eq:epmu}), we obtain $\epsilon_{2r}=-\epsilon_{1r}=1.3$, $\mu_{2r}=-\mu_{1r}=1.3$,
$\epsilon_{2l}=1.1$, $\epsilon_{1l}=-1.5$, $\mu_{2l}=1.1$, and $\mu_{1l}=-1.5$, which satisfy the first of Eq.~(\ref{eq:cmp3}).
In Fig.~\ref{fig71}(a) we plot the transmittances $T_{11}$ and $T_{22}$ through the bilayer in the same configuration as in Fig.~\ref{fig71b}(a)
and confirm that the omnidirectional total transmission of RCP waves occurs when $\nu$ is zero.
For a small value of $\nu$ equal to $10^{-5}$, the transmittance remains equal to one only at $\theta<40.54^\circ$ for RCP waves.
There is no omnidirectional total transmission for LCP waves and the transmittance $T_{22}$ depends very weakly on $\nu$.
In Fig.~\ref{fig71}(b) we consider the transmittances in the same configuration as in Fig.~\ref{fig71b}(b) and
find that the omnidirectional total transmission of LCP waves occurs when $\nu$ is zero.

In the present case, having the impedance matching in the whole space is a necessary condition to have the omnidirectional total transmission
for RCP or LCP waves.
In Fig.~\ref{fig19} we consider the same bilayer as in Figs.~\ref{fig71b}(a) and \ref{fig71}(a) and assume that the medium parameters
in the incident region and substrate are $\epsilon_i=\epsilon_t=4$, $\mu_i=\mu_t=1$, and $a_i=a_t=0$.
Then the impedance matching in the whole space is not achieved and the omnidirectional total transmission does not occur.
In Fig.~\ref{fig19}(a) we show that the absorptance $A_2$ is no longer zero due to the coupling between LCP and RCP waves.
In Fig.~\ref{fig19}(b) we confirm that
the transmittance $T_{11}$ is smaller than one at all $\theta$ even when $\nu$ is zero.

\section{Conclusion}
\label{sec5}

In this paper, we have investigated the characteristics of the surface waves excited at the interface between
two different bi-isotropic media.
We have derived an analytical dispersion relation for those waves, using which we have deduced
the conditions under which they are excited between
two Tellegen media and between two chiral media
for all or a wide range of incident angles in ATR experiments on multilayer structures.
We have also obtained the conditions under which
the omnidirectional total transmission occurs through conjugate matched pairs.
We have found that the omnidirectional excitation of surface waves and the omnidirectional total transmission through a conjugate matched pair
are intimately related phenomena and discussed the similarities and differences of the respective conditions.
We have also pointed out that the omnidirectional total transmission discussed here has basically the same physical origin as the super-Klein tunneling occurring
in pseudospin-1 Dirac-type materials.
We have confirmed our predictions with detailed numerical calculations of the absorptance, the transmittance, and the spatial distribution of the electromagnetic fields.

The omnidirectional excitation of surface waves and the omnidirectional total transmission through a conjugate matched pair
extensively discussed in previous sections
require that the chirality index $\gamma$ or the Tellegen parameter $\chi$ takes a fairly large value.
Though there exist no natural materials with such properties, there have been many recent theoretical and experimental studies to construct
artificial metamaterial structures with a very large and tunable value of $\gamma$ \cite{plum,min1,song,min2}. It may be possible to test our theory
using those metamaterials.

It is more difficult to construct artificial metamaterials with a large value of the Tellegen parameter $\chi$.
Recently, it has been pointed out that some topological Dirac materials such as topological insulators and Weyl semimetals
can be considered electromagnetically as a kind of Tellegen medium \cite{13,14,silv,sarma,sono,jafa}.
The $\chi$ value for topological insulators is very small, but it may be possible to obtain strongly
enhanced magnetoelectric effects in Weyl semimetals with tilted Dirac cones.
Further research in that direction is highly desired.

\acknowledgments
This research was supported through a National Research Foundation of Korea Grant (NRF-2020R1A2C1007655) funded by the Korean Government.


\begin{thebibliography}{99}

\bibitem{polo} J. A. Polo, Jr., T. Mackay, and A. Lakhtakia, {\it Electromagnetic Surface Waves: A Modern
Perspective} (Elsevier, Waltham, MA, 2013).
\bibitem{taka} O. Takayama, A. A. Bogdanov, and A. V. Lavrinenko, J. Phys.: Condens. Matter {\bf 29}, 463001 (2017).


\bibitem{mayer} K. M. Mayer and J. H. Hafner, Chem. Rev. {\bf 111}, 3828 (2011).
\bibitem {petry} E. Petryayeva and U. J. Krull, Anal. Chim. Acta {\bf 706}, 8 (2011).
\bibitem{amen} V. Amendola, R. Pilot, M. Frasconi, O. M. Marag\`o, and M. A. Iat\`i, J. Phys.: Condens. Matter {\bf 29}, 203002 (2017).


\bibitem{maier} S. A. Maier, {\it Plasmonics: Fundamentals and Applications} (Springer, New York, 2007).
\bibitem{mis} M. I. Stockman {\it et al.}, J. Opt. {\bf 20}, 043001 (2018).

\bibitem{raether} H. Raether, {\it Surface Plasmons on Smooth and Rough Surfaces and on Gratings} (Springer-Verlag, Berlin, 1988).


\bibitem{taka2} O. Takayama, L. Crasovan, D. Artigas, and L. Torner, Phys. Rev. Lett. {\bf 102}, 043903 (2009).
\bibitem{nari} E. E. Narimanov, Phys. Rev. A {\bf 98}, 013818 (2018).

\bibitem{xue} C.-H. Xue, H.-T. Jiang, H. Lu, G.-Q. Du, and H. Chen, Opt. Lett. {\bf 38}, 959 (2013).
\bibitem{8} K. J. Lee, J. W. Wu, and K. Kim, Opt. Express {\bf 21}, 28817 (2013).
\bibitem{abf} B. Augui\'e, M. C. Fuertes, P. C. Angelom\'e, N. L. Abdala, G. J. A. A. Soler Illia,
and A. Fainstein, ACS Photonics {\bf 1}, 775 (2014).

\bibitem{12} G. Mi and V. Van, Opt. Lett. {\bf 39}, 2028 (2014).
\bibitem{nah} M. Naheed, M. Faryad, and T. G. Mackay, J. Opt. Soc. Am. B {\bf 36}, F1 (2019).

\bibitem{9} K. Park, B. J. Lee, C. Fu, and Z. M. Zhang,
J. Opt. Soc. Am. B {\bf 22}, 1016 (2005).
\bibitem{doc} J. A. Dockrey, S. A. R. Horsley, I. R. Hooper, J. R. Sambles, and A. P. Hibbins, Sci. Rep. {\bf 6}, 22018 (2016).

\bibitem{10} K. Kim, Opt. Express {\bf 16}, 13354 (2008).
\bibitem{shv} A. B. Shvartsburg, N. V. Silin, and Y. G. Nesterov, Phys. Status Solidi B {\bf 256}, 1800697 (2019).

\bibitem{gal} V. M. Galynsky, A. N. Furs, and L. M. Barkovsky, J. Phys. A:
 Math. Gen. {\bf 37}, 5083 (2004).

\bibitem{ksp1} K. Kim, D. K. Phung, F. Rotermund, and H. Lim, Opt. Express {\bf 16}, 15506 (2008).
\bibitem{ksp2} K. Kim, J. Korean Phys. Soc. {\bf 67}, 2092 (2015).
\bibitem{ksp3} S. Kim and K. Kim, Opt. Express {\bf 25}, 31816 (2017).



\bibitem{sw} S. Kim and K. Kim, Opt. Express {\bf 24}, 15882 (2016).

\bibitem{1} I. V. Lindell, A. H. Sihvola, S. A. Tretyakov, and A. J. Viitanen, {\it Electromagnetic Waves
in Chiral and Bi-Isotropic Media} (Artech House, Boston, 1994).
\bibitem{2} J. Lekner, Pure Appl. Opt. {\bf 5}, 417 (1996).

\bibitem{kly} V. I. Klyatskin, Prog. Opt. {\bf 33}, 1 (1994).
\bibitem{kim1} K. Kim, Phys. Rev. B {\bf 58}, 6153 (1998).
\bibitem{kim3} K. Kim, D.-H. Lee, and H. Lim, Europhys. Lett. {\bf 69}, 207 (2005).
\bibitem{kim4} K. Kim, D. K. Phung, F. Rotermund, and H. Lim, Opt. Express {\bf 16}, 1150 (2008).
\bibitem{15} S. Kim and K. Kim, J. Opt. {\bf 18}, 065605 (2016).

\bibitem{alu} A. Al\`u and N. Engheta, IEEE Trans. Antennas Propag. {\bf 51}, 2558 (2003).

\bibitem{shen} R. Shen, L. B. Shao, B. Wang, and D. Y. Xing,
Phys. Rev. B {\bf 81}, 041410(R) (2010).
\bibitem{urban} D. F. Urban, D. Bercioux, M. Wimmer, and W. Hausler,
Phys. Rev. B {\bf 84}, 115136 (2011).
\bibitem{fang0} A. Fang, Z. Q. Zhang, S. G. Louie, and C. T. Chan,
Phys. Rev. B {\bf 93}, 035422 (2016).
\bibitem{bo1} Y. Betancur-Ocampo, G. Cordourier-Maruri, V. Gupta, and R. de Coss,
Phys. Rev. B {\bf 96}, 024304 (2017).
\bibitem{kim_rip} K. Kim, Results Phys. {\bf 12}, 1391 (2019).
\bibitem{kimd2} S. Kim and K. Kim, Phys. Rev. B {\bf 100}, 104201 (2019).

\bibitem{kats2} M. I. Katsnelson, K. S. Novoselov, and A. K. Geim, Nat. Phys. {\bf 2}, 620 (2006).
\bibitem{been} C. W. J. Beenakker, Rev. Mod. Phys. {\bf 80}, 1337 (2008).
\bibitem{nicol} E. Illes and E. J. Nicol, Phys. Rev. B {\bf 95}, 235432 (2017).
\bibitem{kimd1} S. Kim and K. Kim, Phys. Rev. B {\bf 99}, 014205 (2019).

\bibitem{flach} D. Leykam, A. Andreanov, and S. Flach, Adv. Phys. X {\bf 3}, 1473052 (2018).

\bibitem{plum} E. Plum, J. Zhou, J. Dong, V. A. Fedotov, T. Koschny, C. M. Soukoulis, and N. I. Zheludev,
Phys. Rev. B {\bf 79}, 035407 (2009).
\bibitem{min1} T.-T. Kim, S. S. Oh, H.-S. Park, R. Zhao, S.-H. Kim, W. Choi,
B. Min, and O. Hess, Sci. Rep. {\bf 4}, 5864 (2014).
\bibitem{song} K. Song, Z. Su, M. Wang, S. Silva, K. Bhattarai, C. Ding,
Y. Liu, C. Luo, X. Zhao, and J. Zhou, Sci. Rep. {\bf 7}, 10730 (2017).
\bibitem{min2} H. S. Park, J. Park, J. Son, Y. Kim, H. Cho, J. Shin,
W. Jeon, and B. Min, Adv. Optical Mater. {\bf 7}, 1801729 (2019).

\bibitem{13} A. Karch, Phys. Rev. B {\bf 83}, 245432 (2011).
\bibitem{14} L. L. Li and W. Xu, Appl. Phys. Lett. {\bf 104}, 111603 (2014).
\bibitem{silv} F. R. Prud\^encio and M. G. Silveirinha, Phys. Rev. A {\bf 93}, 043846 (2016).
\bibitem{sarma} J. Hofmann and S. Das Sarma, Phys. Rev. B {\bf 93}, 241402(R) (2016).
\bibitem{sono} K. Sonowal, A. Singh, and A. Agarwal, Phys. Rev. B {\bf 100}, 085436 (2019).
\bibitem{jafa} Z. Jalali-Mola and S. A. Jafari, Phys. Rev. B {\bf 100}, 205413 (2019).


\end{thebibliography}

\end{document}